\documentclass[11pt,a4paper,twoside]{article}

\usepackage[utf8]{inputenc}
\usepackage[T1]{fontenc}
\usepackage[english]{babel} 
\usepackage{geometry}
\usepackage{amsmath, amsfonts, amssymb, amsthm}
\usepackage{bm} 

\usepackage{graphicx}
\usepackage{booktabs}
\usepackage{caption}
\usepackage{subcaption}

\usepackage{authblk} 

\usepackage[round]{natbib} 
\usepackage[breaklinks=true]{hyperref}
\hypersetup{
    colorlinks=true,
    linkcolor=blue,
    filecolor=magenta,      
    urlcolor=blue,
    citecolor=blue,
    pdftitle={Titre de votre article},
    pdfauthor={Votre Nom}
}
\usepackage{orcidlink} 

 \usepackage{fancyhdr}

\fancypagestyle{premierepage}{
    \fancyhf{} 

    \chead{%
        \parbox{\textwidth}{%
            \centering \scriptsize \textit{This is the peer-reviewed version of the article accepted for publication in the Quarterly Journal of the Royal 
             Meteorological Society, with the final published version available at \url{https://doi.org/10.1002/qj.70177}}%
            
        }%
    }

    \cfoot{\thepage}
}

\usepackage{algpseudocode}
\usepackage{algorithm}
\usepackage{algorithmicx}
\usepackage{algcompatible}
\usepackage{algorithm}

\usepackage{float} 

\makeatletter
\newcommand\blfootnote[1]{%
  \begingroup
  \renewcommand\thefootnote{}\footnote{#1}%
  \addtocounter{footnote}{-1}%
  \endgroup
}
\makeatletter

\title{\textbf{A Structurally Localized Ensemble Kalman Filtering  Approach}}

\author{Boujemaa Ait-El-Fquih and Ibrahim Hoteit}

\affil{\small King Abdullah University of Science and Technology (KAUST), Thuwal, Saudi Arabia}
\date{}

\begin{document}

\maketitle
\thispagestyle{premierepage} 

\blfootnote{Correspondence to: boujemaa.aitelfquih@gmail.com, ibrahim.hoteit@kaust.edu.sa}


\begin{abstract}
\noindent
 State-of-the-art ensemble Kalman filtering (EnKF) algorithms require incorporating localization techniques  to cope with the rank deficiency and the inherited  spurious correlations in their error covariance matrices. Localization techniques are mostly {\it ad-hoc},    based on some distances between the state and observation variables,  requiring demanding manual tuning. This work introduces a new ensemble filtering approach, which is inherently localized,  avoiding the need for   any auxiliary localization technique. Instead of explicitly applying localization on  ensembles, the idea is to first localize the continuous analysis probability density function (pdf) before ensemble sampling. The localization of the analysis pdf  is performed through an approximation by a product of independent marginal pdfs corresponding to small partitions of the state vector, using the variational Bayesian optimization. These marginals are then sampled following stochastic EnKF and deterministic ensemble transform Kalman filtering (ETKF) procedures, using ensembles larger than the partitions' size. The resulting  filters  involve the same forecast steps as their standard EnKF and ETKF counterparts but different  analysis steps, iteratively   adjusting the   EnKF   and ETKF updates of each partition based on the  ensemble means of the other partitions. Numerical experiments are conducted with the Lorenz-96 model under different scenarios to demonstrate the potential  of the proposed filters. The new filters' performances are comparable to those of the EnKF and ETKF with already tuned localization, both in terms of computational burden and estimation accuracy.  
\end{abstract}

\vspace{1em}
\noindent\textbf{Keywords:} Data assimilation; EnKF; ETKF; Localization; Variational Bayes.

\section{Introduction}
\label{ref-sec-1} 

Bayesian filtering methods, also known as sequential data assimilation methods, follow a probabilistic framework in which the underlying state estimation problem is split into successive cycles of alternating forecast and analysis steps \citep{kunsch-chap3-2001, hoteit-et-al-chapter-2018}.     
 The forecast step computes the forecast probability density function (pdf) of the current state given past observations, by integrating the previous analysis pdf with the state model. The forecast pdf is then updated in the analysis step with the incoming observation to obtain the analysis pdf of the state given all observations up to the current time instant.  
  Such forecast and analysis pdfs can be, in theory,  exactly evaluated using the famous Kalman filter (KF) in the particular case of  linear-Gaussian state-space systems \citep{Kalman1,andersonmoore}. In practice, however, the use of the KF is limited to relatively small systems, to avoid  a prohibitive cost  for  the computation and storage of  large error covariance matrices \citep[e.g., ][]{aitelfquih-et-hoteit-2022-i3etsp}. This makes it impossible the use of KF in realistic large-scale geophysical applications, which  further involves nonlinear systems. 
  
In  nonlinear but small systems, good  Monte Carlo approximations for  the forecast and analysis distributions can be obtained using the particle filter (PF), a  popular numerical filtering method which provides  mathematically sound  schemes with asymptotic convergence properties and ease of implementation \cite[e.g.,][]{gordon-salmond-smith, doucet-sequentialMC}. The PF applicability is limited to small systems because of the exponential relationship between the state dimension and the  number of particles that need to be generated to efficiently sample  the state-space,   the so-called  curse of dimensionality issue \citep{crisan-doucet-2002-tsp, Snyder_obstaclesto-2008, HoteitP2008}.   
  Some  improved variants of PF have been introduced in an attempt to address  both nonlinear and large-dimensional aspects \citep[e.g., ][]{ades-vanleeuwen-2013, aBeskos-et-al-2014-aap,    septier-et-peters-2015, aitelfquih-et-hoteit-2016-sp-vbnlg, jPoterjoy-2016-mwr,  rPtthast-et-al-2019-mwr,  cDiMauro-et-al-2022-wrr}. In parallel, a distinct class of Bayesian inference methods known as Particle Flow Filters, which do not rely on importance weighting, has demonstrated applicability to high-dimensional geophysical problems \citep[e.g., ][]{mPulido-and-vanLeeuwen-2019-jcp, ccHu-et-vanLeeuwen-2021-qjrms, ccHu-et-al-2024-mwr}. However, despite their better robustness to  the curse of dimensionality compared to the original  PF, these are not up to standard yet in operational geophysical applications.

To date, the golden standard in many geophysical applications is still the ensemble KF (EnKF), thanks to its non-intrusive formulation and ease of implementation, remarkable robustness and effectiveness, and reasonable computational requirements \citep[see e.g.,][and references therein]{reichle-et-al-2002,  edwards-et-al-2015-arms,   andersonn-hoar-et-al-2009, hoteit-et-al-2015, hoteit-et-al-chapter-2018}. The EnKF    provides  Gaussian-based Monte Carlo approximations of the state forecast and analysis distributions by  ensembles of  random realizations, customarily called ensemble members \citep{evensen-1994,  evensen-book-2006, hoteit-et-al-2015}.  In an EnKF cycle, the forecast step integrates the  (analysis) ensemble  forward with the dynamical state model to obtain a forecast ensemble of the new state,   
 and the  analysis step computes an analysis ensemble of the same state from the resulting forecast ensemble using a KF-like update based on the incoming observation and the observation  error model \citep{hoteit-et-al-chapter-2018}. 
 
 In its original form (hereafter referred to as the stochastic EnKF, SEnKF), it  perturbs  the observations stochastically before applying the KF update to each ensemble member 
 \cite[e.g.,][]{burgers-et-al-98, paper-houtekamer-Mitchell-mwr-1998, hoteit-et-al-2015}.
 The perturbation of the observations guarantees an asymptotic  matching of the covariance of the sampled  analysis ensemble to that of the theoretical KF  \citep{burgers-et-al-98}. However,  despite the remarkable performances shown by the stochastic SEnKF in many applications, it underestimates the  covariance   when the  ensemble size is smaller than the rank of the observation noise \citep{hoteit-et-al-2015}. 
   This phenomenon can be either i) mitigated through for instance incorporating in the SEnKF an {\sl improved sampling scheme} as in  \cite{evensen-2004-oceanDynamics, evensen-book-2009} or a so-called second-order resampling scheme as in \cite{hoteit-et-al-2015}, or ii)  fully avoided using a deterministic (square root) EnKF formulation.  
    Indeed, the  deterministic EnKFs avoid perturbing  the observations and sample the  analysis ensemble from an update of  the mean  and a square-root form of the covariance of the forecast ensemble exactly as in the KF, based on the actual observation \citep[e.g.,][]{pham-monthweaRev-2001, bishop-et-al-monthweaRev-2001, hoteit-et-al-2002, TippettA2003, hunt-et-al-physica-2007, hoteit-et-al-2015}.     
    Several deterministic EnKFs are available in the literature, mainly differing in the form of the analysis covariance square-root and the way it is  matched.  The most  popular with publicly available codes being  the ensemble transform KF 
(ETKF) \citep{bishop-et-al-monthweaRev-2001, Wang-et0al-20224-mwr, hunt-et-al-physica-2007}, 
the singular evolutive interpolated KF \citep{pham-monthweaRev-2001, hoteit-et-al-2002}, 
and the ensemble adjustment KF \citep{paper-qnderson-2001-mwr, andersonn-hoar-et-al-2009}.

 The good performances shown by all stochastic and deterministic  EnKFs in realistic applications  would not be possible without accounting, in their implementation, for   the inevitable   sampling and systematic errors \citep{whitaker-et-al-2008, houtekamer-et-Mitchell-2005}. 
 Sampling  errors stem from the use of small ensembles (compared to the - large - state dimension) in order to limit the  computational cost. Systematic errors   encompass the uncertainties caused by   other  imperfections, namely, the poor knowledge of state  and observation model error statistics, the inadequacy between the nonlinear character of the models and the Gaussian assumption made to derive the filters, etc \citep[e.g.,][and references therein]{inflationPaper-2020}.

 Omitting sampling and systematic errors limits the representativeness of the covariance of the filter  forecast (background) ensemble; this results in   a deficient rank, underestimated  diagonal  variances,  and spurious off-diagonals (cross-variances). This significantly limits the ability of the filter to fit the observations and to produce meaningful  state estimates \citep{whitaker-hamill-2002, furrer-et-Bengeston-2007, hamill-et-all-2009, houtekamer-zhang-2016-mwr, hoteit-et-al-chapter-2018}. 
 Incorporating covariance inflation \citep{paper-qnderson-2001-mwr} and   localization \citep{paper-houtekamer-Mitchell-mwr-1998, hamill-et-al-2001-mwr}  techniques into the EnKFs remain  the most popular among the many approaches  that have been proposed to account for these errors and mitigate their effects.  
Inflation counteracts the problem of underestimation of forecast 
variance by artificially increasing the ensemble spread, either by a multiplicative factor \cite[e.g.][]{AndersonA1999, anderson-tellus-2007,  paper-miyoshi-2011-mwr}, or an additive  factor \citep[e.g. ][]{paper-houtekamer-Mitchell-mwr-2000, paper-whitaker-hamill-mwr-2012},  among others. 
  Localization imposes a decay on the cross-covariance terms with spatial distance, which helps mitigating the issues related to rank deficiency and overestimation of cross-covariances  
   \citep{paper-houtekamer-Mitchell-mwr-1998, hamill-et-al-2001-mwr}.

 Most localization techniques are based on some distances between the  locations of state variables and  observations, the most standard of which are {\sl the local analysis} \citep{paper-houtekamer-Mitchell-mwr-1998} and {\sl covariance localization}  \citep{hamill-et-al-2001-mwr} techniques.  
  Local analysis updates each of the  state variables using only the observations located in its neighbourhood, while the covariance localization  applies a Schur product on the forecast covariance matrix  using a tapering matrix \citep{GaspariC1999}. 
   Other  physical distance-based localization methods have been proposed by for e.g.   \cite{houtekamer-mitchell-2001-mwr, yanhuiZhang-oliver-2010-mgeos, fertig-et-al-2007-tellusA, bishop-hodyss-2007-qjrms}.  
   Further works have also been  conducted to deal with the situations where some or all of the state and/or observation variables cannot be associated with physical locations \citep[e.g.][]{ bocquet-2016-qjrms, bishop-hodyss-2007-qjrms, anderson-physicaD-2007, delaChevrotiere-harlim-2017-mwr, lu-et-al-20018-spe}.

 All   EnKF localization schemes  explicitly apply on  ensembles  (discrete distributions),  i.e., one first derives the  ensemble sampler of the continuous analysis pdf (i.e., the EnKF) then  incorporates localization within it. 
 Here, we  reverse this order  by first localizing the continuous analysis pdf then deriving  ensemble samplers of the resulting localized pdf. 
  The localization of the analysis pdf   consists of approximating it with a product of separable (independent) marginals pdfs using the (functional) variational Bayesian (VB) optimization, i.e., according to the  Kullback-Leibler divergence (KLD) minimization criterion \citep{livresmidlquin, blei-et-al--2017, urozayev-et-al--2022}. This amounts to removing  posterior interdependencies between  partitions of  the state vector. We then sample the resulting marginal analysis densities according to SEnKF- and ETKF-like analysis steps, but any other sampling scheme could be also used. 
    If the ensemble size is larger than  the partitions' size, the proposed ensemble schemes should work without incorporating  any additional localization scheme. 
   
 The proposed  partitioned  SEnKF (pSEnKF) and ETKF (pETKF) share the same forecast step as the standard SEnKF and ETKF, and have different  analysis steps operating  partition-wise in an iterative way.  More specifically, once classical SEnKF (resp.  classical ETKF) update is  applied for all partitions;  the resulting ensemble (resp. mean ensemble) of each partition is then iteratively adjusted/corrected based on the ensemble means of the other partitions. Such an adjustment  compensates in some way for  the posterior independence imposed between the state partitions. 

 The concept of state partitioning using the VB approach has been already adopted in the linear-Gaussian framework,  
  first in  \citep{aitelfquih-et-hoteit-2015-sp-vblg} where an element-wise mean-field (diagonal posterior covariance) approximation to the KF has been proposed, then in \citep{aitelfquih-et-hoteit-2022-i3etsp} where the element-wise approximation is relaxed to allow for partition-wise mean-field (block-diagonal posterior covariance) approximation. The present work is an extension of this prior work to the more general nonlinear ensemble-like framework.

    The remainder of the paper is organized as follows. Section  \ref{ref-sec-2}  formulates  the  Bayesian filtering problem and reviews  the standard SEnKF and ETKF. 
   Section  \ref{ref-sec-3}  introduces the proposed  localization approach of the continuous analysis pdf. Section  \ref{ref-sec-4}  focuses  on its SEnKF- and ETKF-like sampling, and provides a thorough discussion on the resulting new localized filters. 
   Section \ref{ref-sec-5}  presents and analyses the results of  numerical  experiments that were conducted with the  Lorenz-96 model, before concluding with a summary and general discussion in Section \ref{ref-sec-6}.

\section{Classical Ensemble Kalman Filtering: SEnKF and ETKF}  
\label{ref-sec-2} 

\subsection{Problem formulation}
Let ${\bf x} = {\{ {\bf x}_n  \}}_{n=0}^N$ and ${\bf y} = {\{ {\bf y}_n\}}_{n=0}^N$, with $\mathbf{x}_n \in \mathbb{R}^{d_x}$ and ${\bf y}_n \in \mathbb{R}^{d_y}$, denote a discrete-time (unknown) state process and an observation process, respectively. In  data assimilation, 
 these processes are typically described  according to a  state-space system of the form:   
  \begin{equation} 
\label{DynSys} 
\left\{ 
\begin{array}{ccc} 
	\mathbf{x}_{n} & = & {\cal M}_{n-1}\left( \mathbf{x}_{n-1}\right) + {\bf u}_{n-1} , \\  
	\mathbf{y}_n & = & {\bf H}_n \mathbf{x}_n + {\bf v}_n  ;  
\end{array} 
\right.  
\end{equation} 
  ${\cal M}_{n-1}(.)$  being a (possibly nonlinear) dynamical  operator   integrating the   state   of the system   from  time  $t_{n-1}$ to $t_{n}$, and  ${\bf H}_n$  an   observational  operator at time $t_n$, assumed  linear here for simplicity.  The case of a nonlinear ${\bf H}_n$ could be treated as usual in EnKFs \citep[e.g., ][]{liu-aitelfquih-hoteit-2015}.   
   The  state noise process, ${\bf u} = {\{{\bf u}_n\}}_{n}$, and  the observation noise process, ${\bf v} = {\{{\bf v}_n\}}_{n}$, are assumed to  be   independent, jointly independent and independent of the initial state, ${\bf x}_0$. The noises  ${\bf u}_n$ and ${\bf v}_n$ are further assumed   Gaussian  with zero-means and covariances,  ${\bf Q}_n$ and ${\bf R}_n$, respectively.

The filtering problem refers to the (online) inference of the state ${\bf x}_n$  from the available observations, ${\cal Y}_n = \{{\bf y}_0, \cdots, {\bf y}_n \}$,  and aims at seeking for the analysis pdf, $p_n({\bf x}_n)$, i.e., the pdf of ${\bf x}_n$  conditioned on ${\cal Y}_n$.   When a point estimate is needed, the posterior mean (PM), i.e., the mean of $p_n({\bf x}_n)$, is often chosen as it is the optimum solution according to several criteria including the mean-squared error (MSE) minimization \citep{van-trees-book-1968}.   

The aforementioned   independence assumptions enables a {\sl recursive} calculation of   $p_n({\bf x}_n)$, driven by the state  transition pdf, 
 \begin{equation}  
\label{eq-trans-x-N-v0} 
p({\bf x}_n |{\bf x}_{n-1})   =  {\cal N}_{{\bf x}_n} ({\cal M}_{n-1}({\bf x}_{n-1}) , {\bf Q}_{n-1}) ,  
\end{equation} 
 and the likelihood pdf\footnote{It is worth recalling that  the term ``likelihood'' corresponds to $p({\bf y}_n|{\bf x}_n)$ when ${\bf y}_n$ is a 
fixed value, i.e., the value of the conditional pdf $p({\bf y}_n|{\bf x}_n)$ at a given point ${\bf y}_n$. However, in this paper,  we tolerate, as commonly done, some flexibility  
 and attribute ``likelihood'' to $p({\bf y}_n|{\bf x}_n)$ whether or not ${\bf y}_n$ is fixed.},  
\begin{equation} 
\label{eq-trans-y-N-v0} 
p({\bf y}_n |{\bf x}_{n})  =  {\cal N}_{{\bf y}_n} ({\bf H}_{n}{\bf x}_{n} , {\bf R}_{n}) ;     
\end{equation}    
  ${\cal N}_{{\bf x}}({\bf m}, {\bf C})$ denotes a Gaussian pdf\footnote{The Gaussian forms  (\ref{eq-trans-x-N-v0}) and   (\ref{eq-trans-y-N-v0}) are immediate results of the Gaussian assumptions made on the noises ${\bf u}_{n-1}$ and ${\bf v}_n$, respectively.} of argument ${\bf x}$ and parameters $({\bf m} , {\bf C})$  \citep[e.g., ][]{kunsch-chap3-2001, ieeetrsp-TMC}. 
   An $(n-1,n)$  assimilation  cycle (recursion)  proceeds typically\footnote{A different  filtering approach, involving the one-step-ahead smoothing pdf, can be found in \cite{Chapter-2022} and references therein.} with a  forecast  step which uses the  transition pdf 
   to obtain the forecast pdf, $p_{n-1}({\bf x}_n)$, from the previous  analysis pdf as,    
\begin{equation}
\label{prediction-step}
p_{n-1}({\bf x}_n)  =  \int \! p({\bf x}_{n} | {\bf x}_{n-1}) p_{n-1}({\bf x}_{n-1}) d{\bf x}_{n-1} ,    
\end{equation} 
followed by an analysis step which uses the likelihood to obtain the analysis pdf of the current state following Bayes'  rule, 
\begin{equation} 
\label{filtering-step}
p_n({\bf x}_n) \propto   p({\bf y}_{n} | {\bf x}_{n}) p_{n-1}({\bf x}_{n}) . 
\end{equation}

\subsection{Ensemble Gaussian filtering:  SEnKF and ETKF} 

The  SEnKF and   ETKF are Gaussian-based Monte Carlo implementations of the generic filtering algorithm (\ref{prediction-step})-(\ref{filtering-step}), 
introduced to avoid the linearization of the models and to efficiently handle large-dimensional systems \citep{hoteit-et-al-chapter-2018}. They have been demonstrated to provide robust  PM estimates  even when they are implemented with small ensembles \citep{hoteit-et-al-2015}.

 Both filters involve the same  forecast step, as a Monte Carlo implementation of Eq.  (\ref{prediction-step}),  producing a  state forecast ensemble, ${\{ {\bf f}_{n}^{(m)} \}}_{m=1}^{M}$, after  propagating  the previous analysis ensemble, ${\{ {\bf a}_{n-1}^{(m)} \}}_{m=1}^{M}$,  by the state model, i.e.,   
 \begin{equation}  
\label{eq-forecast-pf-x}
{\bf f}_n^{(m)}  =  {\cal M}_{n-1} ({\bf a}_{n-1}^{(m)}) + {\bf u}_{n-1}^{(m)} , \quad m=1, \cdots, M ;     
\end{equation}  
  ${\bf u}_{n-1}^{(m)}$ being a sample of the Gaussian model error pdf ${\cal N}({\bf 0} , {\bf Q}_{n-1})$.

Given the forecast ensemble and  a new observation ${\bf y}_n$, the SEnKF and ETKF analysis  steps are  then derived under the  assumption of Gaussian\footnote{Note that since in system  (\ref{DynSys}) the  observation model is linear with a Gaussian noise, one only needs to assume the forecast pdf $p_{n-1}({\bf x}_n)$ Gaussian, as it  implies that $p({\bf x}_n,{\bf y}_n|{\cal Y}_{n-1})$ is Gaussian.} 
$p({\bf x}_n,{\bf y}_n|{\cal Y}_{n-1})$, to restrict the problem to a Kalman-like linear update.  
 SEnKF and ETKF have been  respectively classified as  stochastic and deterministic samplers of the same analysis density, $p_n({\bf x}_n)$,  based on the generic update step  (\ref{filtering-step}).

\subsubsection{The (stochastic) SEnKF analysis step}  
\label{sec-standard-SEnKF}  
  
 The SEnKF analysis step uses perturbations ${\bf y}_n^{(m)}$ of the  observation by adding random observational errors ${\bf v}_n^{(m)} \sim {\cal N}({\bf 0}, {\bf R}_n)$, i.e., ${\bf y}_n^{(m)} = {\bf y}_n + {\bf v}_n^{(m)}$, to  update the forecast members according to the KF update step to obtain the analysis ensemble members:  
 \begin{equation}
\label{eq-JSEnKF-xAThA} 
{\bf a}_{n}^{(m)} = {\bf f}_{n}^{(m)}  +  {\bf K}_n    ( {\bf y}_n^{(m)}  -   {\bf H}_n {\bf f}_n^{(m)} ) , \quad m=1, \cdots,  M.   
\end{equation} 
  The Kalman gain ${\bf K}_n$, which is a Monte Carlo approximation of  ${\rm cov}[{\bf x}_n,{\bf z}_n|{\cal Y}_{n-1}] \times \left( {\rm cov}[{\bf z}_n |{\cal Y}_{n-1}] \right.$ $\left. + {\rm cov}[{\bf v}_n] \right)^{-1}$ with  ${\bf z}_n = {\bf H}_n {\bf x}_n$, can be expressed as:  
\begin{equation}
\label{eq-kalman-gain-SEnKF} 
{\bf K}_n =    {\bf S}_{ {\bf f}_{n}} \widetilde{\bf S}_{ {\bf f}_{n}}^T \left[ \widetilde{\bf S}_{ {\bf f}_{n}} \widetilde{\bf S}_{ {\bf f}_{n}}^T  +   {\bf R}_n \right]^{-1},   
\end{equation} 
 where ${\bf S}_{{\bf f}_n}$ denotes the forecast perturbation matrix (i.e., a  square-root of the forecast sample error covariance) and   $\widetilde{\bf S}_{ {\bf f}_{n}} = {\bf H}_n {\bf S}_{ {\bf f}_{n}}$.   
  
   It is important to highlight a theoretical distinction regarding the stochastic update mechanism. While the standard formulation  presented above involves perturbing the observation vector \citep{burgers-et-al-98}, 
   a rigorous derivation   of the stochastic filter from the Bayesian standpoint reveals that the random perturbations should structurally be applied to  ${\bf H}_n {\bf f}_n^{(m)}$, rather than to the observation  \citep{van-leeuwen-qjrms-2020, Chapter-2022},  which leads to involve $-{\bf v}_n^{{(m)}}$ in Eq. (\ref{eq-JSEnKF-xAThA}) instead of $+{\bf v}_n^{{(m)}}$. Nevertheless, observing that both $+{\bf v}_n^{{(m)}}$ and $-{\bf v}_n^{{(m)}}$ are samples of the same {\it symmetric} Gaussian distribution, ${\cal N}({\bf 0}, {\bf R}_n)$, we tolerate,  as customarily done,  some flexibility in the present work and use $+{\bf v}_n^{{(m)}}$ instead.

\subsubsection{The (deterministic) ETKF analysis step} 
\label{sec-standard-ETKF}

Unlike the SEnKF analysis step,  the ETKF analysis step is deterministic in the sense that it directly uses the observation to update  the mean, $\hat{\bf f}_n$, of the forecast members and their perturbation matrix, ${\bf S}_{{\bf f}_n}$, based on the KF update step.  
   The resulting  analysis mean, $\hat{\bf a}_n$, and analysis perturbation matrix, ${\bf S}_{{\bf a}_n}$ (i.e., a square-root of the analysis sample error covariance), are then used to sample the analysis ensemble deterministically following a  matching procedure of the first two moments of the KF solution   \citep[see e.g., ][]{bishop-et-al-monthweaRev-2001,  TippettA2003, hoteit-et-al-2015,  raboudi-et-al-2019-mwr-hybridETKF}.  Formally: 
\begin{itemize}
\item  
 $\hat{\bf a}_n$ results from a KF update of $\hat{\bf f}_{n}$ based on ${\bf y}_n$ as, 
\begin{equation}
\label{xa-mean-etkf} 
\hat{\bf a}_n = \hat{\bf f}_{n}  +  {\bf K}_n    ( {\bf y}_n  -  {\bf H}_n \hat{\bf f}_n ) ;     
\end{equation} 
\item   ${\bf S}_{{\bf a}_n}$ is obtained from a transformation of ${\bf S}_{{\bf f}_n}$ according to the   KF update  of the covariance as, 
\begin{equation}
\label{Sxna-etkf} 
{\bf S}_{{\bf a}_n} = {\bf S}_{{\bf f}_n} {\bf T}_n {\boldsymbol\Omega}_n;   
\end{equation} 
 ${\bf T}_n$ being a square-root of $\mathbb{I}_M + \tilde{\bf S}_{{\bf f}_n}^T {\bf R}_n^{-1} \tilde{\bf S}_{{\bf f}_n}$,  
  and     ${\boldsymbol\Omega}_n$ is a centring matrix guaranteeing the second moment matching (i.e., it ensures that ${\bf S}_{{\bf a}_n} {\bf S}_{{\bf a}_n}^T$ matches the theoretical expression of the analysis covariance of the KF).  Expressions of ${\bf T}_n$ and ${\boldsymbol\Omega}_n$ can be found  e.g., in \cite{bishop-et-al-monthweaRev-2001, pham-monthweaRev-2001, Wang-et0al-20224-mwr}. 

\item  The analysis members are finally sampled as,  
\begin{equation}
\label{eq-JETKF-xAThA} 
{\bf a}_{n}^{(m)} = \hat{\bf a}_{n}  +  \sqrt{M -1} {\left( {\bf S}_{{\bf a}_n} \right)}_m    ; \quad m=1, \cdots, M, 
\end{equation}  
 with ${\left( {\bf S}_{{\bf a}_n} \right)}_m$ denotes the $m$-{\sl th} column of ${\bf S}_{{\bf a}_n}$. 
\end{itemize}

\section{Localization of the theoretical analysis pdf}  
\label{ref-sec-3} 

Let the state vector, ${\bf x}_n$, be split into $K$ partitions, ${\bf x}_{n}^1$, ${\bf x}_{n}^2$, $\cdots$, ${\bf x}_{n}^K$ of the same dimension,  $d_p = d_x/K$. In the more general case where the partitions ${\{{\bf x}_n^k\}}_{k}$ have 
different dimensions ${\{d_{p^k}\}}_{k}$, the proposed schemes remain valid by only replacing $K d_p$ with $\sum_{k=1}^K d_{p^k}$.   
   The analysis pdf,  $p_n({\bf x}_n)$, can be localized by dropping dependencies between  its $K$ marginals.  Specifically, this can be achieved by approximating $p_n({\bf x}_n)$ with  a product of separable marginal pdfs\footnote{The same notation, $\pi(.)$,  will be used for the pdf of  the whole  vector, ${\bf x}_n$, and of those of  partitions, ${\bf x}_{n}^k$},  
\begin{eqnarray} 
 \nonumber 
p_n({\bf x}_n) & \approx &   \pi_n({\bf x}_n) \\ 
\label{eq-sep-VBq} 
& = & \pi_n({\bf x}_{n}^1)  \pi_n({\bf x}_{n}^2) \cdots \pi_n({\bf x}_{n}^K).   
\end{eqnarray}   
 
  We adopt the VB approach and seek for the best approximation $\pi_n({\bf x}_n)$, in the sense of KLD minimization or equivalently   negative free energy (NFE) maximization   \citep[e.g., ][and references therein]{fraysse-rodet-2014}, i.e., 
\begin{equation}
\label{eq-NFE-max}
\pi_n({\bf x}_{n})  =  \operatornamewithlimits{arg \, max}_{q \in \Omega} \,  \mathbb{E}_{q({\bf x}_n)} \!\! \left[ \ln \frac{p({\bf x}_n,{\bf y}_n|{\cal Y}_{n-1})}{q({\bf x}_n)} \right] , 
\end{equation} 
 where $\Omega$ denotes a space of separable pdfs satisfying $q({\bf x}_n) =  \prod_{k=1}^K q({\bf x}_{n}^k)$, and  $\mathbb{E}_{q({\bf x}_n)}[.]$ the expected value with respect to (w.r.t.) $q({\bf x}_n)$.  
 Applying   variational calculus to  maximization (\ref{eq-NFE-max}) readily results in a solution $\pi_n({\bf x}_{n})$ whose marginals satisfy: 
\begin{equation}
\label{eq-theor-vb-approx} 
\pi_n({\bf x}_{n}^k)  \propto   {\cal L}_{{\bf y}_n}({\bf x}_{n}^k) \times  {\pi}_{n-1}({\bf x}_{n}^k), 
\end{equation}  
  with,  
\begin{eqnarray}
\label{vb-lik-theo} 
{\cal L}_{{\bf y}_n}({\bf x}_{n}^k) & \propto &   \exp   \left(\mathbb{E}_{\pi_n({\bf x}_{n}^{k^-})} [\ln p({\bf y}_n | {\bf x}_{n})] \right), \\
\label{vb-prior-theo} 
{\pi}_{n-1}({\bf x}_{n}^k) & \propto &    \exp  \left(\mathbb{E}_{\pi_n({\bf x}_{n}^{k^-})} [\ln p_{n-1}({\bf x}_{n})] \right),    
\end{eqnarray} 
 where $k=1, \cdots, K$ and ${\bf x}_n^{k^-}$ denotes the complement of ${\bf x}_n^k$ in ${\bf x}_n$ \citep{paperieee-smidl-et-al-2008}.

Eqs. (\ref{eq-theor-vb-approx})-(\ref{vb-prior-theo}) suggest that   the proposed approach imposes   independencies between  
  $\pi_n({\bf x}_n^k)$ and  
  $\pi_n({\bf x}_n^{k-}) = \prod_{j \neq k} \pi_n({\bf x}_n^j)$ through ``an averaging'' of  the log likelihood $\ln p({\bf y}_n|{\bf x}_n)$ and the log forecast $\ln p_{n-1}({\bf x}_n)$ w.r.t. $\pi_n({\bf x}_n^{k-})$.  
   In other words, for each $k=1, \cdots, K$, the partition ${\bf x}_n^{k}$ of the full state, ${\bf x}_n$, is not localized by crudely ignoring the other partitions, ${\bf x}_n^{k-}$, but by ``freezing'' them by means of the aforementioned ``averaging''.        
 On the other hand, one can notice that the $K$ Eqs. (\ref{eq-theor-vb-approx}) have the form of  a Bayesian update step, linking priors  
  ${\pi}_{n-1}({\bf x}_{n}^k)$ with posteriors 
   $\pi_n({\bf x}_{n}^k)$  using likelihoods ${\cal L}_{{\bf y}_n}({\bf x}_{n}^k)$.

    Connection between the expression of the true analysis pdf $p_n({\bf x}_n^k)$ (i.e.,  marginal of $p_n({\bf x}_n)$) and that of its VB estimate, $\pi_n({\bf x}_{n}^k)$, can be then made by rewriting $p_n({\bf x}_n^k)$ in a similar functional form as $\pi_n({\bf x}_{n}^k)$ (Eqs. (\ref{eq-theor-vb-approx})-(\ref{vb-prior-theo})). This is achieved by  applying the logarithm function on both sides of Eq. (\ref{filtering-step}), taking the expectation w.r.t.  $p_n({\bf x}_{n}^{k^-})$, and finally applying the exponential function. This yields:          
\begin{equation}
\label{eq-theor-vb-approx-pnsin} 
p_n({\bf x}_{n}^k)  \propto  \frac{ {\kappa}_{{\bf y}_n}({\bf x}_{n}^k) \times \tilde{p}_{n-1}({\bf x}_{n}^k) }{ 
\exp  \left(\mathbb{E}_{p_n({\bf x}_{n}^{k^-})} [\ln p({\bf x}_{n}^{k-}|{\bf x}_n^k,{\cal Y}_n)] \right) 
}, 
\end{equation}  
  with,  
\begin{eqnarray}
\label{vb-lik-theo-pnsin} 
{\kappa}_{{\bf y}_n}({\bf x}_{n}^k) & \propto &   \exp   \left(\mathbb{E}_{p_n({\bf x}_{n}^{k^-})} [\ln p({\bf y}_n | {\bf x}_{n})] \right), \\
\label{vb-prior-theo-pnsin} 
\tilde{p}_{n-1}({\bf x}_{n}^k) & \propto &    \exp  \left(\mathbb{E}_{p_n({\bf x}_{n}^{k^-})} [\ln p_{n-1}({\bf x}_{n})] \right).     
\end{eqnarray}  
  Now, given that Eqs. (\ref{vb-lik-theo-pnsin})-(\ref{vb-prior-theo-pnsin}) are similar to Eqs. (\ref{vb-lik-theo})-(\ref{vb-prior-theo}), the expression (\ref{eq-theor-vb-approx-pnsin}) of $p_n({\bf x}_{n}^k)$ ends up with an extra term  dependent on ${\bf x}_n^k$,  
  $T_{{\bf y}_n}({\bf x}_n^k) = \exp  \left(\mathbb{E}_{p_n({\bf x}_{n}^{k^-})} [\ln p({\bf x}_{n}^{k-}|{\bf x}_n^k,{\cal Y}_n)] \right)$,    compared to the expression    (\ref{eq-theor-vb-approx}) of $\pi_n({\bf x}_{n}^k)$. 
  In the particular case where the partitions are by essence independent conditionally on ${\cal Y}_n$, this extra term vanishes (as it becomes independent of ${\bf x}_n^k$) and expressions  (\ref{eq-theor-vb-approx}) and (\ref{eq-theor-vb-approx-pnsin}) coincide (i.e., ${\cal L}_{{\bf y}_n}({\bf x}_{n}^k) = {\kappa}_{{\bf y}_n}({\bf x}_{n}^k) $, ${\pi}_{n-1}({\bf x}_{n}^k) = \tilde{p}_{n-1}({\bf x}_{n}^k)$ and $\pi_n({\bf x}_n^k) = p_n({\bf x}_n^k)$).    
  Such a  scenario corresponds to state-space systems that do not involve 
connections between the different partitions (as if one has $K$ independent systems, one 
for each partition).  This is a very restrictive condition, that is not generally fulfilled 
in real-world applications. 
 
 In standard systems (i.e., in which state partitions connect with each other),   
   the VB approach eliminates the term $T_{{\bf y}_n}({\bf x}_n^k)$ (to make expressions  (\ref{eq-theor-vb-approx}) and (\ref{eq-theor-vb-approx-pnsin})  identical) by approximating it with a constant quantity, independent of ${\bf x}_n^k$, which is then  included in  the normalization term of  Bayes' formula (\ref{eq-theor-vb-approx}). 
 One can show that this approximation is performed as  an ``averaging''  w.r.t. $p_n({\bf x}_n^k)$, i.e.,
 \begin{eqnarray} 
\nonumber 
\pi_n({\bf x}_{n}^k)  & \stackrel{ (\ref{eq-theor-vb-approx}) }{\propto}  & \frac{ {\cal L}_{{\bf y}_n}({\bf x}_{n}^k) \times  {\pi}_{n-1}({\bf x}_{n}^k)  }{ 
\int_{\mathbb{R}^{d_p}} {\cal L}_{{\bf y}_n}({\bf x}_{n}^k) \times  {\pi}_{n-1}({\bf x}_{n}^k) d{\bf x}_n^k } , \\  
\label{eq-theor-vb-approx-001}  
  & \propto  & \frac{ {\cal L}_{{\bf y}_n}({\bf x}_{n}^k) \times  {\pi}_{n-1}({\bf x}_{n}^k)  }{ \mathbb{E}_{p_n({\bf x}_n^k)} [\tilde{T}_{{\bf y}_n}({\bf x}_n^k) ]
  }  ;  
\end{eqnarray}  
    $\tilde{T}_{{\bf y}_n}({\bf x}_n^k) = \exp  \left(\mathbb{E}_{\pi_n({\bf x}_{n}^{k^-})} [\ln p({\bf x}_{n}^{k-}|{\bf x}_n^k,{\cal Y}_n)] \right)$.     
    This  means that to provide a separable KLD estimation to the analysis pdf, the VB approach  modifies  the standard (true) Bayesian approach by ``freezing'' the $K$ terms $T_{{\bf y}_n}({\bf x}_n^k)$  by means of ``an averaging'' over ${\bf x}_n^k$. 
   
 Finally, it is worth noting that in a Gaussian framework, the VB approach does not impose block diagonal structure to the true posterior pdf, but rather seeks to find the Gaussian pdf with block diagonal covariance that is closest (in the sense of criterion (\ref{eq-NFE-max})) to the true posterior pdf. As such, even with the separable  approximation (\ref{eq-sep-VBq}) that implies block-diagonal covariance matrix in a  Gaussian framework, the true posterior pdf does not necessarily have a block-diagonal covariance.

\subsection{Calculation of the localized analysis pdfs $\pi_n({\bf x}_n^k)$} 
\label{sub-sec-loacalAnalysisPDF}
 
As discussed above, although  the VB marginal analysis pdfs, $\pi_n({\bf x}_n^k)$, are, by construction, mutually independent, each of them is still  related to a functional form involving the  expectation w.r.t. the  others, $\pi_n({\bf x}_n^{k^-})$. Such a functional relationship compensates for the enforced posterior independencies between the state partitions, a key advantage of the VB approach. 
  However, it  also has the drawback of precluding an  exact evaluation of the  $K$ solutions (\ref{eq-theor-vb-approx}) and approximations should  therefore  be performed.

  One natural way to do so is  to proceed with (coordinate descent-like)  iterations \cite[e.g.,][]{bertsekas-Belmont-1995}. That is, at iteration $i$,  update successively each of the $K$ marginals $\pi_n^{i}({\bf x}_n^k)$ using the newly updated $\pi_n({\bf x}_n^{k^-})$.  
  In that respect, the update (\ref{eq-theor-vb-approx}) at iteration $i$ reads:      
\begin{equation} 
\label{eq-vbIter-classic-partition} 
\pi_n^i({\bf x}_{n}^k)  \propto   {\cal L}_{{\bf y}_n}^i({\bf x}_{n}^k) \times {\pi}_{n-1}^i({\bf x}_{n}^k),    
\end{equation} 
 where ${\cal L}_{{\bf y}_n}^i({\bf x}_{n}^k)$ and ${\pi}_{n-1}^i({\bf x}_{n}^k)$ are respectively given by (\ref{vb-lik-theo}) and (\ref{vb-prior-theo}) using   $\pi_n^{(i,i-1)}({\bf x}_n^{k-}) = \prod_{j=1}^{k-1}  \pi_n^{i}({\bf x}_n^{j}) \times \prod_{j=k+1}^{K}   \pi_n^{i-1}({\bf x}_n^{j})$ in the expectation operator  (recall that  $ \pi_n({\bf x}_n^{k-}) = \prod_{j=1}^{k-1}  \pi_n({\bf x}_n^{j}) \times \prod_{j=k+1}^{K}   \pi_n({\bf x}_n^{j})$).  
 
\subsection{The generic localized filtering algorithm} 
\label{sec-generic-local-filter} 
 
 Here, we present the forecast and analysis steps of the generic localized filtering  algorithm for  System (\ref{DynSys}). 
Denote hereafter  ${\bf H}_n^k$ and ${\bf H}_n^{k-}$ the partitions of  ${\bf H}_n$ corresponding to the positions $k$ and $k-$, respectively, i.e.,   satisfying  ${\bf H}_n {\bf x}_n = {\bf H}_n^k {\bf x}_n^k + {\bf H}_n^{k-} {\bf x}_n^{k-}$.  Let also $(\hat{\bf x}_{n|n-1}, {\bf P}_{n|n-1})$ the first two moments of the forecast pdf ${p}_{n-1}({\bf x}_{n})$, and $({\bf m}_{n|j}^{k,i} , {\bf C}_{n|j}^k)$  the first two moments of $\pi_{j}^{i}({\bf x}_n^k)$ with $j=n-1, n$ (we show in the Appendix   that the covariances ${\bf C}_{n|j}^k$ are invariant over the iterations).  
 \\[0.7em] 
\noindent $\bullet$ {\sl  Forecast step.} \,  $p_{n-1}({\bf x}_n)$ is calculated  from $\pi_{n-1}({\bf x}_{n-1})$ following the standard forecast step, i.e.,  following Eq. (\ref{prediction-step}) using $\pi_{n-1}({\bf x}_{n-1})$ instead of $p_{n-1}({\bf x}_{n-1})$. \\[0.7em]
 $\bullet$  {\sl Analysis step.} \, The resulting forecast pdf,  $p_{n-1}({\bf x}_n)$, is updated based on the newly acquired observation, ${\bf y}_n$, following an iterative procedure to obtain a localized analysis pdf, $\pi_n({\bf x}_n)$. Starting from an initialization $\pi_n^{0}({\bf x}_n) = p_{n-1}({\bf x}_n)$, the $i$-th iteration calculates  the marginals $\pi_n^{i}({\bf x}_n^k)$ successively from $k=1$ to $K$ 
 in three steps: 
\begin{enumerate}
\vspace{-.0cm}
\item A first step computes the VB likelihood ${\cal L}_{{\bf y}_n}^i({\bf x}_{n}^k)$ based on Eq. (\ref{vb-lik-theo}) using $\pi_n^{(i,i-1)}({\bf x}_n^{k-})$ in the expectation operator. One obtains, 
\begin{equation}  
\label{eq-trans-y-N-v0-vb} 
{\cal L}_{{\bf y}_n}^i({\bf x}_{n}^k)  =  {\cal N}_{{\bf y}_n} ({\bf H}_{n}^k {\bf x}_{n}^k + {\bf H}_n^{k-} {\bf m}_{n|n}^{k-,i}, {\bf R}_{n}) ; 
\end{equation} 
${\bf m}_{n|n}^{k-,i} =  ({\bf m}_{n|n}^{1,i}, \cdots, {\bf m}_{n|n}^{k-1,i}, {\bf m}_{n|n}^{k+1,i-1}, \cdots, {\bf m}_{n|n}^{K,i-1})$. 

\item A second step applies ``an adjustment'' to the forecast ${p}_{n-1}({\bf x}_{n})$  according to   Eq. (\ref{vb-prior-theo}) using $\pi_n^{(i,i-1)}({\bf x}_n^{k-})$ in the expectation operator, to obtain ${\pi}_{n-1}^i({\bf x}_{n}^k)$.  Assuming that ${p}_{n-1}({\bf x}_{n})$ is Gaussian,   one obtains,    
\begin{eqnarray}
\label{vb-trans-pdf-xk} 
{\pi}^i_{n-1}({\bf x}_{n}^k)  & =  & {\cal N}_{{\bf x}_n^k} ({\bf m}_{n|n-1}^{k,i} , {\bf C}_{n|n-1}^k) ; \\ 
\label{vb-trans-pdf-xk-mean} 
{\bf m}_{n|n-1}^{k,i}  & =  & \hat{\bf x}_{n|n-1}^k + {\bf G}_n^k ({\bf m}_{n|n}^{k-,i} - \hat{\bf x}_{n|n-1}^{k-}), \\ 
\label{vb-trans-pdf-xk-cov}  
{\bf C}_{n|n-1}^k  & =  & {\bf P}_{n|n-1}^k - {\bf G}_n^k {\bf P}_{n|n-1}^{k-,k},  
\end{eqnarray} 
with ${\bf G}_n^k = {\bf P}_{n|n-1}^{k,k-} ({\bf P}_{n|n-1}^{k-})^+$, where $(.)^+$ denotes  the 
 pseudo-inverse, and the covariances ${\bf P}_{n|n-1}^{k}$ and ${\bf P}_{n|n-1}^{k-}$ and the cross-covariances ${\bf P}_{n|n-1}^{k,k-}$ and ${\bf P}_{n|n-1}^{k-,k}$ are extracted from ${\bf P}_{n|n-1}$ according to the positions $k$ and $k^-$.

\item   The third step combines the Gaussian VB likelihood (\ref{eq-trans-y-N-v0-vb}) with the Gaussian VB forecast (\ref{vb-trans-pdf-xk})-(\ref{vb-trans-pdf-xk-cov}) according to the Bayesian  update  (\ref{eq-vbIter-classic-partition}),  to obtain a Gaussian VB analysis pdf:    
\begin{eqnarray}
\label{vb-trans-pdf-xk-analysis} 
 {\pi}^i_{n}({\bf x}_{n}^k)  & =  & {\cal N}_{{\bf x}_n^k} ({\bf m}_{n|n}^{k,i} , {\bf C}_{n|n}^k) ; \\ 
\label{vb-trans-pdf-xk-mean-analysis} 
  {\bf m}_{n|n}^{k,i}  & =  &  {\bf m}_{n|n-1}^{k,i}  +   {\bf L}_n^k ({\bf y}_n  -  {\bf H}_n^k {\bf m}_{n|n-1}^{k,i} \!\! - \! {\bf H}_n^{k-} {\bf m}_{n|n}^{k-,i}), \\ 
\label{vb-trans-pdf-xk-cov-analysis}  
  {\bf C}_{n|n}^k  & =  & {\bf C}_{n|n-1}^k - {\bf L}_n^k {\bf H}_n^k ({\bf C}_{n|n-1}^k)^T ,    
\end{eqnarray} 
with ${\bf L}_n^k = {\bf C}_{n|n-1}^k ({\bf H}_n^k)^T [{\bf H}_n^k {\bf C}_{n|n-1}^k ({\bf H}_n^k)^T + {\bf R}_n]^{-1}$. 
\end{enumerate}  
At convergence (last iteration, $i = I$), ${\pi}_{n}({\bf x}_{n}^k) = {\pi}^I_n({\bf x}_n^k)$ for all $k$. 

\section{Practical implementation: The partitioned-SEnKF (pSEnKF) and -ETKF (pETKF)}  
\label{ref-sec-4}

Here, we focus on SEnKF- and ETKF-like ensemble implementations of the generic algorithm presented in  Section \ref{sec-generic-local-filter}. For simplicity, we use  the  same notations used in Section \ref{ref-sec-2}  for the ensembles and their  corresponding moments (sample means, sample covariances and perturbation  matrices). 

The filters pSEnKF and pETKF  share the same forecast step as their standard counterparts (Eq. (\ref{eq-forecast-pf-x})). 
  The  derivation of their analysis steps  based on the generic KF-like updates  ((\ref{vb-trans-pdf-xk-mean})-(\ref{vb-trans-pdf-xk-cov}),  (\ref{vb-trans-pdf-xk-mean-analysis})-(\ref{vb-trans-pdf-xk-cov-analysis})) may involve computational challenges related to the gain matrices ${\bf G}_n^{k}$.  
  Indeed, in an ensemble setting, these matrices can be   computed based on the forecast perturbation matrices as,  
\begin{eqnarray} 
\nonumber 
{\bf G}_n^k & \approx & {\bf S}_{{\bf f}_n^k} {\bf S}_{{\bf f}_n^{k-}}^T ({\bf S}_{{\bf f}_n^{k-}} \, {\bf S}_{{\bf f}_n^{k-}}^T)^+ , \\
\label{eq-GaginG-ensembles} 
 & = &  {\bf S}_{{\bf f}_n^k}  \, ({\bf S}_{{\bf f}_n^{k-}})^+ .    
\end{eqnarray}   
 In this localized setting, this requires a beforehand computation of  $K$ $(M \times (d_x - d_p))$-sized  pseudo-inverses, followed by $K$ products (\ref{eq-GaginG-ensembles}), in addition to the need for storing the $K$ computed $(d_p \times (d_x - d_p))$-sized  ${\bf G}_n^k$ matrices  during all iterations, as they are needed in the  update  (\ref{vb-trans-pdf-xk-mean}) at every iteration. This can quickly become cumbersome when dealing with  large-dimensional systems. This is dealt with in this preliminary work by simply ignoring the terms involving ${\bf G}_n^k$ in Eqs. (\ref{vb-trans-pdf-xk-mean})-(\ref{vb-trans-pdf-xk-cov}), reducing them to: 
\begin{equation} 
\label{vb-trans-pdf-xk-mean-approx} 
{\bf m}_{n|n-1}^{k,i}   =   \hat{\bf x}_{n|n-1}^k \quad {\rm and}  \quad 
{\bf C}_{n|n-1}^k   =   {\bf P}_{n|n-1}^k,  
\end{equation} 
which in turns reduces the VB analysis updates (\ref{vb-trans-pdf-xk-mean-analysis})-(\ref{vb-trans-pdf-xk-cov-analysis}) into, 
\begin{eqnarray} 
\nonumber 
  {\bf m}_{n|n}^{k,i}  & =  &  \hat{\bf x}_{n|n-1}^{k}  +   {\bf L}_n^k ({\bf y}_n  -  {\bf H}_n^k \hat{\bf x}_{n|n-1}^{k}  -  {\bf H}_n^{k-} {\bf m}_{n|n}^{k-,i}), \\ 
\label{vb-trans-pdf-xk-mean-analysis-approx}  
 & =  &  \underbrace{ \hat{\bf x}_{n|n-1}^{k}  +   {\bf L}_n^k ({\bf y}_n  -  {\bf H}_n^k \hat{\bf x}_{n|n-1}^{k})}_{{\Theta}_n^k \, \rm (classical \, KF)}  -  \underbrace{ {\bf L}_n^k {\bf H}_n^{k-} {\bf m}_{n|n}^{k-,i}}_{\rm adjustment}, \\  
\label{vb-trans-pdf-xk-cov-analysis-approx}  
  {\bf C}_{n|n}^k  & =  & {\bf P}_{n|n-1}^k - {\bf L}_n^k {\bf H}_n^k ({\bf P}_{n|n-1}^k)^T .     
\end{eqnarray} 

Under the Gaussian assumption on $p_{n-1}({\bf x}_n)$,  ignoring  ${\bf G}_n^k$ in Eqs. (\ref{vb-trans-pdf-xk-mean})-(\ref{vb-trans-pdf-xk-cov}) amounts to ignoring the dependencies between ${\bf x}_n^k$ and ${\bf x}_{n}^{k-}$ conditionally on ${\cal Y}_{n-1}$, i.e., to assuming,  
\begin{equation}
\label{eq-assumption-sepForecastPDF}
p({\bf x}_n^k|{\bf x}_{n}^{k-}, {\cal Y}_{n-1}) = p_{n-1}({\bf x}_n^k). 
\end{equation}   
   Noticing that this assumption leads to ignore the forecast cross-covariance across partitions, it can be interpreted as an implicit localization of the forecast distribution.   

Eq. (\ref{vb-trans-pdf-xk-mean-analysis-approx}) suggests that at a given  iteration, the VB analysis mean of a state partition, ${\bf x}_n^k$, is none  other than  a classical KF analysis mean, $\Theta_n^k$, adjusted with a shift by a  linear  combination of the most recent values of  the VB analysis means of the remaining  partitions, ${\bf x}_n^{k-}$.  
  Furthermore, $\Theta_n^k$ is computed only once, as it does not depend on iterations; this means that the iterative update (\ref{vb-trans-pdf-xk-mean-analysis-approx}) of the VB KF analysis mean, ${\bf m}_{n|n}^{k}$, is simply an iterative adjustment of the  classical KF analysis mean, $\Theta_n^k$, based on a  linear  combination of the most recent  ${\bf m}_{n|n}^{k-}$.

\subsection{The pSEnKF analysis step}   

The pSEnKF analysis step is an ensemble form of the  iterative VB  analysis step  (\ref{vb-trans-pdf-xk-mean-analysis-approx}). That is, for each $k$-th partition,  a classical SEnKF update is first performed  
 to obtain  an ensemble ${ \{ \boldsymbol\theta_n^{k,(m)} \}}_{m=1}^M$, which is then adjusted  iteratively  
 with a shift by a linear combination of the most recent sample means $\hat{\bf a}_{n}^{k-}$, 
 to obtain  the VB analysis ensemble of interest, ${\{ {\bf a}_n^{k,(m)} \}}_{m=1}^M$. 
 
By analogy with the expression of $\Theta_n^k$ in Eq. (\ref{vb-trans-pdf-xk-mean-analysis-approx}) (and with Eqs. (\ref{eq-JSEnKF-xAThA})-(\ref{eq-kalman-gain-SEnKF})), the classical SEnKF update reads  for $k=1, \cdots, K$,   
\begin{eqnarray}
\label{vb-etkf-gain} 
\tilde{\bf L}_n^k   & = & {\bf S}_{{\bf f}_n^k} \tilde{\bf S}_{{\bf f}_n^k}^T    [\tilde{\bf S}_{{\bf f}_n^k} \tilde{\bf S}_{{\bf f}_n^k}^T + {\bf R}_n]^{-1},         \\ 
\label{vb-classical-SEnKF} 
\boldsymbol\theta_n^{k,(m)} & =   &  {\bf f}_{n}^{k,(m)}  +   \tilde{\bf L}_n^k ({\bf y}_n^{(m)}  -  {\bf H}_n^k {\bf f}_{n}^{k,(m)}) ,     
\end{eqnarray} 
with  $\tilde{\bf S}_{{\bf f}_n^k} = {\bf H}_n^k {\bf S}_{{\bf f}_n^k}$, ${\bf y}_n^{(m)} = {\bf y}_n + {\bf v}_n^{(m)}$ and  ${\bf v}_n^{(m)} \sim {\cal N}({\bf 0}, {\bf R}_n)$. 
   The iterations are then triggered. At iteration $i$, the VB analysis ensemble, ${\{ {\bf a}_n^{k,(m),i} \}}_{m=1}^M$, is computed via an adjustment of the obtained   ${\{ \boldsymbol\theta_n^{k,(m)} \}}_{m=1}^M$ with a shift by $- \tilde{\bf L}_n^k {\bf H}_n^{k-} \hat{\bf a}_{n}^{k-,i}$, where the sample mean $\hat{\bf a}_{n}^{k-,i}$ is given as,     $\hat{\bf a}_{n}^{k-,i} =  (\hat{\bf a}_{n}^{1,i}, \cdots, \hat{\bf a}_{n}^{k-1,i}, \hat{\bf a}_{n}^{k+1,i-1}, \cdots, \hat{\bf a}_{n}^{K,i-1})$.    
  Specifically, at iteration $i$, one has for $k=1, \cdots, K$:      
\begin{equation} 
\label{SEnKF-Updated-vbmean} 
{\bf a}_n^{k,(m),i} = \boldsymbol{\theta}_{n}^{k,(m)} - \tilde{\bf L}_n^k {\bf H}_n^{k-} \hat{\bf a}_{n}^{k-,i}  ; \;\; m=1, \cdots, M. 
\end{equation} 
 At  convergence (last iteration $I$), ${\{ {\bf a}_n^{k,(m)} \}}_{m=1}^M =  {\{ {\bf a}_n^{k,(m),I} \}}_{m=1}^M$ and  $\hat{\bf a}_n^{k} = \hat{\bf a}_n^{k,I}$ for all $k$. Convergence is assumed to be reached if, for instance, a  predefined maximum number of iterations, $I$, is exceeded, or if an error based on the difference between two successive $\hat{\bf a}_n = (\hat{\bf a}_n^1, \cdots, \hat{\bf a}_n^K)$ (i.e., based on $\hat{\bf a}_n^{i} - \hat{\bf a}_n^{i-1}$)   become less than a given threshold.

An $(n-1,n)$ assimilation cycle of the pSEnKF is outlined in Algorithm \ref{algorithme-pSEnKF}.   The computational load is further discussed in Section  \ref{subsec-discussions}.   

 \begin{algorithm}
  \caption{\!\!{\bf : } \, pSEnKF} 
  \label{algorithme-pSEnKF}
  \begin{algorithmic} [1]  
   
  \Statex  \hspace{-.6cm}  $\bullet$ {\sl \underline{Forecast:}} \, Sample the ensemble ${\{ {\bf f}_n^{(m)} \}}_{m=1}^M$ using Eq. (\ref{eq-forecast-pf-x}),   then inflate it before computing  $\hat{\bf f}_n$ and  ${\bf S}_{{\bf f}_n}$.     
  \vspace{.2cm}  
  
\Statex  \hspace{-.6cm} $\bullet$ {\sl  \underline{Analysis:}} 
   \vspace{.08cm}

   \Statex \hspace{-.25cm} {\sl  
    $1)$ \hspace{-.001cm} \underline{Classical SEnKF:}} \, For $k=1, \cdots, K$,  compute $\tilde{\bf L}_n^k$ using \Statex  \hspace{.2cm} Eq.  
       (\ref{vb-etkf-gain}) then ${\{ \boldsymbol\theta_n^{k,(m)} \}}_{m=1}^M$ using Eq. (\ref{vb-classical-SEnKF}).

 \vspace{.08cm} 
   
   \Statex \hspace{-.25cm} {\sl  
    $2)$ \hspace{-.01cm} \underline{Iterative adjustment:}} \, 
     \vspace{.08cm} 
   
   \Statex  \hspace{.1cm}   
     $\triangleright$  Initialize $\hat{\bf a}_{n} = \hat{\bf f}_n$ and  ${\bf h}_n = {\bf H}_n \hat{\bf f}_n$. 
     \vspace{.08cm} 
   
   \Statex  \hspace{.1cm}  
     $\triangleright$ {\bf While} {\it not converged} {\bf do}:   
     for  $k = 1, \cdots, K$,  
   \begin{eqnarray}
   \nonumber 
   {\bf h}_n & = & {\bf h}_n - {\bf H}_n^k \hat{\bf a}_n^{k}  \\
   \nonumber 
   {\bf a}_n^{k,(m)} & = & \boldsymbol{\theta}_{n}^{k,(m)} - \tilde{\bf L}_n^k  {\bf h}_n; \quad m=1, \cdots, M  \\ 
   \nonumber 
   \hat{\bf a}_n^k & = & \frac{1}{M} \sum_{m=1}^M {\bf a}_n^{k,(m)} \\ 
 \nonumber 
 {\bf h}_n & = & {\bf h}_n + {\bf H}_n^k \hat{\bf a}_n^{k} 
   \end{eqnarray}    
  
  \vspace{-.15cm} 
 \Statex  \hspace{.58cm}   {\bf End}

\end{algorithmic}  
\end{algorithm}

\subsection{The pETKF analysis step}  

The pETKF analysis step is derived based on the KF-like Eqs. (\ref{vb-trans-pdf-xk-mean-analysis-approx})-(\ref{vb-trans-pdf-xk-cov-analysis-approx})  following the principle of ETKF  (Section \ref{sec-standard-ETKF}), i.e.,  ``the moments'' $\hat{\bf a}_n$ and ${\bf S}_{{\bf a}_n}$ are first computed  partition-wise  then used to sample the members ${\bf a}_n^{(m)}$.

 The analysis means, $\hat{\bf a}_n$, are computed following the iterative KF  adjustment procedure (\ref{vb-trans-pdf-xk-mean-analysis-approx}). That is, for each $k$-th partition,  a classical KF update is first performed to obtain  an analysis mean, $\hat{\boldsymbol\theta}_n^{k}$, which is then adjusted  iteratively, based on  a linear combination of the most recent $\hat{\bf a}_{n}^{k-}$, to obtain  the VB  analysis mean of interest, $\hat{\bf a}_n^{k}$. 

By analogy with the expression of $\Theta_n^k$ in Eq. (\ref{vb-trans-pdf-xk-mean-analysis-approx}), the classical KF update reads  for $k=1, \cdots, K$,   
\begin{equation}  
\label{vb-vb-etkf-mean}   
  \hat{\boldsymbol{\theta}}_{n}^{k}  =     \hat{\bf f}_{n}^{k}  +   \tilde{\bf L}_n^k ({\bf y}_n  -  {\bf H}_n^k \hat{\bf f}_{n}^{k}).        
\end{equation}  
At iteration $i$, the VB analysis mean, $\hat{\bf a}_n^{k,i}$, is computed via an adjustment of the obtained   $\hat{\boldsymbol{\theta}}_{n}^{k}$ with a shift by $- \tilde{\bf L}_n^k {\bf H}_n^{k-} \hat{\bf a}_{n}^{k-,i}$:      
\begin{equation} 
\label{etkf-Updated-vbmean} 
\hat{\bf a}_n^{k,i} = \hat{\boldsymbol{\theta}}_{n}^{k} - \tilde{\bf L}_n^k {\bf H}_n^{k-} \hat{\bf a}_{n}^{k-,i}.  
\end{equation} 
 The iteration loop  is terminated following the same stopping criterion discussed in  pSEnKF. 
  Assuming  convergence is reached after $I$ iterations, one takes   $\hat{\bf a}_n^{k} = \hat{\bf a}_n^{k,I}$ for all $k$.

  Square-roots of the sample analysis covariances are  computed based on Eq. (\ref{vb-trans-pdf-xk-cov-analysis-approx}) in a similar way to Eq. (\ref{Sxna-etkf}) as, 
\begin{equation}
\label{Sxna-etkf-vb} 
{\bf S}_{{\bf a}_n^k} = {\bf S}_{{\bf f}_n^k} {\bf T}_n^k {\boldsymbol\Omega}_n,    
\end{equation} 
 where the transformation matrices, ${\bf T}_n^k$, are  here given by square roots of $\mathbb{I}_M +  \tilde{\bf S}_{{\bf f}_n^k}^T {\bf R}_n^{-1} \tilde{\bf S}_{{\bf f}_n^k}$. 
 
 The resulting ${\{ \hat{\bf a}_n^k \}}_{k=1}^{K}$ and ${\{ {\bf S}_{{\bf a}_n^k} \}}_{k=1}^{K}$ are then  concatenated column-wise to respectively form $\hat{\bf a}_n$ and ${\bf S}_{{\bf a}_n}$, based on which the analysis members, ${\bf a}_n^{(m)}$, are sampled as in Eq. (\ref{eq-JETKF-xAThA}).

An $(n-1,n)$ assimilation cycle of the pETKF is summarized in Algorithm \ref{algorithme-pETKF}, and the computational load is discussed in Section  \ref{subsec-discussions}.      

 \begin{algorithm}
  \caption{\!\!{\bf : } \, pETKF} 
  \label{algorithme-pETKF}
  \begin{algorithmic} [1]  
   
  \Statex  \hspace{-.6cm}  $\bullet$ {\sl \underline{Forecast:}} \, Sample the ensemble ${\{ {\bf f}_n^{(m)} \}}_{m=1}^M$ using Eq. (\ref{eq-forecast-pf-x}),   then inflate it before computing  $\hat{\bf f}_n$ and  ${\bf S}_{{\bf f}_n}$.     
  \vspace{.2cm}  
  
\Statex  \hspace{-.6cm} $\bullet$ {\sl  \underline{Analysis:}} 
   \vspace{.08cm}

   \Statex \hspace{-.25cm} {\sl 
    $1)$ \hspace{-.001cm} \underline{Mean:}} \,  
   
   \vspace{.08cm} 
   
  \Statex  \hspace{.1cm}  {\sl $\triangleright$  \hspace{-.01cm} \underline{Classical KF:}} \, For $k=1, \cdots, K$,  Compute $\tilde{\bf L}_n^k$ using Eq. 
 
      \Statex  \hspace{.4cm} (\ref{vb-etkf-gain}), then $\hat{\boldsymbol{\theta}}_{n}^{k}$ using Eq. (\ref{vb-vb-etkf-mean}).    
   
    \vspace{.08cm} 
 \Statex  \hspace{.1cm}  {\sl $\triangleright$  \hspace{-.01cm} \underline{Iterative Adjustment:}} \,     
     \vspace{.08cm} 
   
   \Statex  \hspace{.37cm} $\diamond$ Initialize $\hat{\bf a}_{n} = \hat{\bf f}_n$ and  ${\bf h}_n = {\bf H}_n \hat{\bf f}_n$. 
     \vspace{.08cm} 
     
    \Statex  \hspace{.37cm} $\diamond$  {\bf While} {\it not converged} {\bf do}:   
     for  $k = 1, \cdots, K$,  
   \begin{eqnarray}
   \nonumber 
   {\bf h}_n & = & {\bf h}_n - {\bf H}_n^k \hat{\bf a}_n^{k}  \\
   \nonumber 
   \hat{\bf a}_n^k & = & \hat{\boldsymbol{\theta}}_{n}^{k} - \tilde{\bf L}_n^k  {\bf h}_n  \\ 
 \nonumber 
 {\bf h}_n & = & {\bf h}_n + {\bf H}_n^k \hat{\bf a}_n^{k} 
   \end{eqnarray}  
  
  \vspace{-.15cm} 
 \Statex  \hspace{.8cm}   {\bf End}
  
  \vspace{.08cm}   
   
   \Statex  \hspace{-.25cm} {\sl  
    $2)$ \underline{Covariance square-root:}} \, For $k=1, \cdots, K$, compute ${\bf S}_{{\bf a}_n^k}$ 
    
     \Statex \hspace{.08cm} from ${\bf S}_{{\bf f}_n^k}$ using  Eq. (\ref{Sxna-etkf-vb}).

   \vspace{.08cm}    
   
  \Statex \hspace{-.2cm} {\sl $3)$ \underline{Members:}} \,  Sample the ensemble ${\{ {\bf a}_n^{(m)} \}}_{m=1}^M$ following  Eq. 
  \Statex \hspace{.08cm}  (\ref{eq-JETKF-xAThA}) based on the resulting  $\hat{\bf a}_n$ and ${\bf S}_{{\bf a}_n}$.      

\end{algorithmic}
\end{algorithm}

 \subsection{Discussion} 
 \label{subsec-discussions}

The  proposed generic filtering scheme   in  Section  \ref{sec-generic-local-filter} shares the same forecast step as the standard generic filtering  algorithm (Eq. (\ref{prediction-step})), but involves a different analysis step that operates partition-wise by   splitting the state ${\bf x}_n$ into low-dimensional sub-states, ${\bf x}_n^{k}$, $k=1, \cdots, K$, following the VB approach.   
  Such an analysis step is endowed with an iterative procedure;  each iteration,  $i$, consists of  $K$ successive Bayesian update steps linking respectively the $K$ state-partitions ${\bf x}_n^k$  
  to the observation ${\bf y}_n$, in order to compute the VB analysis pdfs $\pi_n^i({\bf x}_n^k)$. 

 In the particular case where the observation model is linear, such $K$  Bayesian updates reduce to $K$ KF-like updates computing means and covariances of the  pdfs $\pi_n^i({\bf x}_n^k)$. The covariances of $\pi_n^i({\bf x}_n^k)$ 
 do not depend on the iterations and are thus computed  once, outside of  the  iteration loop, using the classical KF update applied on the  models ${\bf y}_n = {\bf H}_n^k {\bf x}_n^k + {\bf v}_n$.  
 The mean of each of the $\pi_n^i({\bf x}_n^k)$ varies with the iterations as  an iterative adjustment of the classical KF analysis mean (i.e., obtained from   applying KF on the model ${\bf y}_n = {\bf H}_n^k {\bf x}_n^k + {\bf v}_n$), 
 with a shift by  
 a  linear  combination of the most recent values of the means that correspond to the  remaining partitions, ${\bf x}_{n}^{k-}$.  
 
The  proposed pSEnKF and pETKF algorithms are   
  ensemble variants  of the afore-described generic filtering algorithm,    under the Gaussian  assumption on the forecast pdf and the independence assumption  (\ref{eq-assumption-sepForecastPDF}). They share the same forecast step with the standard SEnKF and ETKF filters, but involve different analysis steps that are respectively stochastic and  deterministic ensemble forms of the afore-described iterative generic analysis step. 
  More specifically, the pSEnKF analysis step consists of an iterative SEnKF  adjustment procedure based on linear combinations of sample analysis means   (i.e., adjustments of ensembles generated by the standard SEnKF analysis),  whereas the pETKF analysis step employs exactly the above generic iterative KF adjustment procedure 
  before (deterministically) generating the  analysis ensembles.  
  
  Splitting the state vector into partitions could be seen as an  intrinsic localization, negating the need for (external) auxiliary  localization, most notably when the ensemble size, $M$, is larger than the  partitions' size, $d_p$. In fact, unlike the  SEnKF and ETKF,  in  which the localization is applied on  discrete distributions, i.e., on  (forecast or analysis) ensembles, the proposed approach starts by applying  localization  on the (continuous) analysis density, based on the VB approach,  then sampling the resulting localized pdf by pSEnKF and pETKF algorithms without equipping them with any (external) localization, subject that $M > d_p$.  
  
It is also worth noting that the aforementioned (in Section   \ref{sub-sec-loacalAnalysisPDF}) functional relationship that  links  each  VB analysis pdf $\pi_n^i({\bf x}_n^k)$ to expectation w.r.t. $\pi_n^{(i,i-1)}({\bf x}_n^{k^-})$, appears in the pSEnKF and pETKF analysis  steps under the form of a correction term, $-\tilde{\bf L}_n^k {\bf H}_n^{k-} \hat{\bf a}_n^{k-,i}$ (i.e., Eqs. (\ref{SEnKF-Updated-vbmean}) and (\ref{etkf-Updated-vbmean})). This means that contrary to what one might expect, the partitioning that is applied on the (continuous) analysis pdf using the VB approach does not make the  analysis pSEnKF and pETKF ensembles completely  independent between the partitions, as these are still linked via their sample means, i.e., ${\{ {\bf a}_n^{k,(m),i} \}}_m$ are linked to $\hat{\bf a}_n^{k-,i}$.    
   More precisely, this coupling acts at the mean estimate level but does not reintroduce full cross-partition (analysis) covariances.  As a result, when the true posterior exhibits strong dependencies across partition boundaries (i.e., when  boundaries split strongly coupled neighbors), the analysis update may preserve these couplings reasonably well in the ensemble mean (through the iterative mean correction), while  cross-boundary correlations within the analysis ensemble are not preserved,  resulting in  physical imbalances.  
     Nevertheless, this behavior, which is analogous to the imbalances observed in  standard local analysis schemes \citep[e.g.,][]{hunt-et-al-physica-2007}, does not compromise the filter stability. Indeed, the physical balances are  restored during the subsequent forecast step, where the model dynamics act as a smoother,  re-establishing the cross-partition correlations before the next assimilation cycle.

  The convergence of the proposed iterative VB algorithms to a local minimum of  {\sl theoretical} KLD  is not guaranteed, notably owing to the underlying Gaussian and posterior independence assumptions on which these algorithms are founded   \citep[e.g. ][]{stochParamPaper-2023,  livresmidlquin, m-sato-2011, m-a-chappell-et-al-2009,  blei-et-al--2017, pollution-paper-2019}. However, convergence  towards a local minimum of the  ``{\sl estimated}'' KLD, i.e., approximated based on the  ensembles,  is guaranteed by construction  \citep[e.g.][]{pollution-paper-2019}.

Characterizing  the accuracy of VB-like approximations is not easy   
  as this depends on several factors, including the number of partitions, $K$, and the strength  of the posterior  dependencies between these  partitions \citep[e.g.][]{jordan-et-al-1999,  jordan-et-jordan-2000, variationalTutorial-2001, livresmidlquin}.    
  Instead, one can intuitively examine the question of when the accuracy of VB approximation is  likely to be reasonable and when one may expect it to fail.  
   Clearly, when the state partitions are weakly ({\sl a posteriori}) dependent, the VB estimate $\pi_n({\bf x}_n) = \prod_{k=1}^K \pi_n({\bf x}_n^k)$ of $p_n({\bf x}_n)$ should be reasonable.   
      In contrast, when these partitions become strongly dependent,  the ``assumption''  of conditional independence becomes  strong and this  may jeopardize the VB estimation.

The VB estimation should (theoretically) improve with decreased number of  partitions $K$, or  equivalently with increased dimension $d_p$.  
 However, in practice, since the pSEnKF and pETKF algorithms (without auxiliary localization) would be  useful only when $d_p$ does not exceed the ensemble size, $M$, 
  the rule of thumb would be  to  choose, for a given $M$,  a $d_p$ that is as large as possible but not exceeding  $M$.

The proposed pSEnKF and pETKF practically have similar algorithms  as their  respective standard counterparts, with the {\sl Iterative Adjustment} as the sole extra component (see Algorithms $1$ and $2$). The Iterative Adjustment step has a computational cost that scales linearly with the state dimension, $d_x$,  approximately equal to ${\cal C}_{\rm pSEnKF}^{\rm extra} = \left[ 2 d_y + (6 d_y + 2M) I \right] d_x$ flops\footnote{floating operations, see e.g., \cite{aitelfquih-et-hoteit-2015-sp-vblg}.} in the pSEnKF and   ${\cal C}_{\rm pETKF}^{\rm extra} =  \left[ 2 d_y + 6 d_y I \right] d_x$ flops in the pETKF.  
 The iterations' number, $I$, is typically much smaller than $d_x$ and $d_y$, a key property of the VB-type optimization algorithms \citep[e.g.,][]{m-sato-2011}. Thus,   ${\cal C}_{\rm pSEnKF}^{\rm extra}$ and  ${\cal C}_{\rm pETKF}^{\rm extra}$ should be insignificant  in large-scale geophysical applications, compared to the costs of the  SEnKF and ETKF analysis steps ($\sim {\cal O}(d_y^2 d_x)$ in linear observation models) and  forecast step.

 It is finally worth noting that the proposed iterative schemes are VB coordinate-descent procedures (specifically, cyclic coordinate ascent on the negative free energy) with analytical Kalman-like updates, rather than numerical linear solvers (e.g., Krylov subspace methods). Consequently, standard physical preconditioners are not directly applicable to these probabilistic updates. However, in  high-dimensional or non-Gaussian scenarios, acceleration techniques may become useful, for instance, preconditioned Conjugate Gradient for solving large-scale innovation-covariance inversions, or natural-gradient methods \citep{martens-2020-jmlr} for optimization-based non-Gaussian updates.

\section{Numerical experiments}    
\label{ref-sec-5} 
Numerical experiments are conducted with the  nonlinear Lorenz-96  (L96) model  \citep{LorenzE1998}   
  to assess the behavior of the proposed pSEnKF and pETKF,  and to   evaluate their performances against the standard SEnKF and ETKF using localization.   
  L96  model simulates the time evolution of an atmospheric quantity based on differential equations: 
\begin{equation} 
\frac{dx(j,t)}{dt}   =   x(j-1,t)[x(j+1,t)-x(j-2,t)]-x(j,t)+ F , 
\label{eq-model-L96}
\end{equation} 
where  $F = 8$ denotes the external forcing term, and $x(j,t)$, with $j=1, \cdots, d_x = 40$,      the $j$-th  element of the state at  time  $t$; boundaries conditions are periodic (i.e., $x(-1,t)=x(d_x-1,t)$,  $x(0,t)=x(d_x,t)$, and $x(1,t)=x(d_x+1,t)$).

The model  (\ref{eq-model-L96}) is numerically integrated using the  fourth-order Runge-Kutta discretization scheme with a time step  $\delta_m = 0.05$, equivalent to six hours in real time.  The model is assumed  perfect,     
  and integrated forward to simulate a trajectory of $114600$ time steps, starting from an initial state which  elements (state variables) are set to $F$, except the $20$-th one, which is set to   $(1 + 0.001) F$. 
  The first $100000$ time steps of the resulting trajectory are  discarded as a spin-up period, and the remaining $14600$ time steps (i.e., $10$  years in model time)  are  considered  as the  reference state trajectory, to be later estimated by the filters based on observations extracted from those reference states. 
  The observations are  extracted by adding to the states  a Gaussian noise with zero mean and a  covariance,  ${\bf R} = r \times \mathbb{I}_{d_x}$,  with $r = 0.6$, corresponding to a signal-to-noise-rate (SNR) of $15$ dB, i.e.,  
 \begin{equation}
\nonumber 
r = \frac{\sum_{n=0}^{T-1} \| {\bf H}_n {\bf x}_n \|^2}{T d_y} \times 10^{-{\rm SNR}/10};  
\end{equation} 
  $T$ being  the total number of model time steps \citep[e.g.][]{aitelfquih-et-hoteit-2015-sp-vblg}.

      We start by presenting the results of an experiment aiming at demonstrating the relevance of 
   the  proposed pSEnKF and pETKF with $M = 20$  members,   in a full observational scenario (i.e., all state variables are observed) and a setting where  the observations are assimilated  every four model time steps (i.e., with an observations' time step,  $\delta_o = 4 \, \delta_m$).  
  In both filters, the initial forecast  ensemble is  generated from a Gaussian distribution centred around  the mean  of the model trajectory, and having $3 \mathbb{I}_{d_x}$ as covariance.  Their (iterative) analysis steps  split the state  into $K= 4$ partitions (of dimension $d_p = 10$), and   assume convergence  to be reached when the relative squared misfit norm of the full state analysis  estimate,  ${\rm RSMN}^i = \| \hat{\bf a}_n^{i} - \hat{\bf a}_n^{i-1} \|^2 / \| \hat{\bf a}_n^{i-1} \|^2 $,    
  drops below $10^{-10}$   (this is equivalent to monitor convergence using NFE  (\ref{eq_nfe-at-iteration-i})-(\ref{eq-nfe-phiTerms}), as in the proposed filters only the analysis means vary over iterations whereas the covariances are kept fixed).         
   Further both filters  use inflation with a   factor, of $1.1$, chosen based on  trial-and-error tuning.   
  
   To reduce statistical fluctuations, each experiment is repeated independently for $50$ times, with different realizations of   observational noises and initial ensembles.  
   The average of the $50$ runs is taken as the final result.

\begin{figure}[H] 
\centering 
\centerline{\includegraphics[width=10cm]{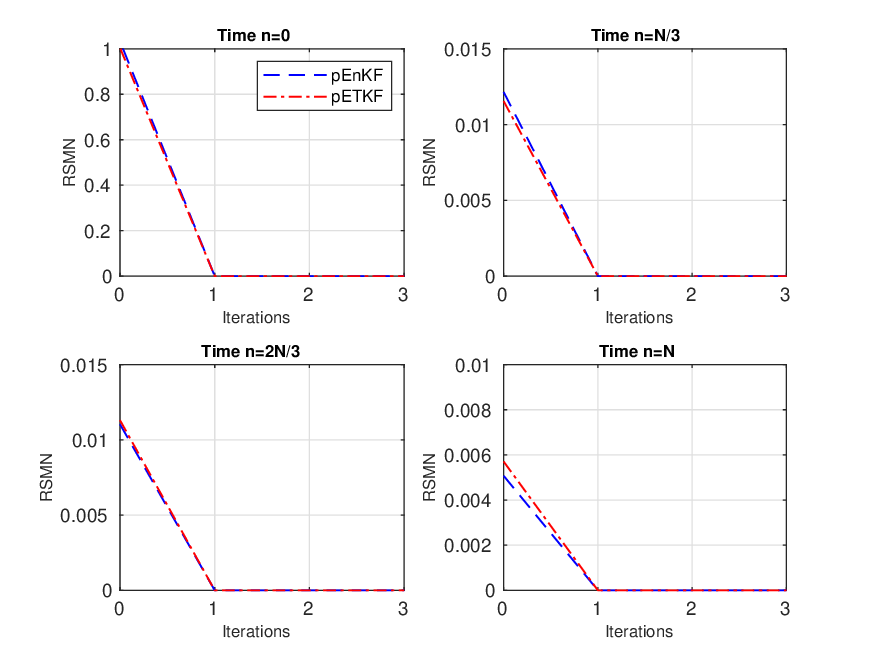}}  
\caption{Evolution of the RSMN of the state analysis estimate as function of the iterations'   number after every ${\rm N/3}$ assimilation cycles, with ${\rm N}$ being their total number.  
}\label{figure_1}
\end{figure}

   Fig. \ref{figure_1} displays  the evolution over  iterations of  ${\rm RSMN}$ at assimilation cycles $n = 0$, $N  /3$, $2N /3$ and $N$, with $N$ being the total number of assimilation cycles.  As can be seen, both algorithms achieve  convergence after $2$ iterations only, consistent with the   well-known fast convergence of VB-like filtering algorithms   \citep[e.g.][]{m-sato-2011}.

\begin{figure}[H] 
\centering 
\centerline{\includegraphics[width=10cm]{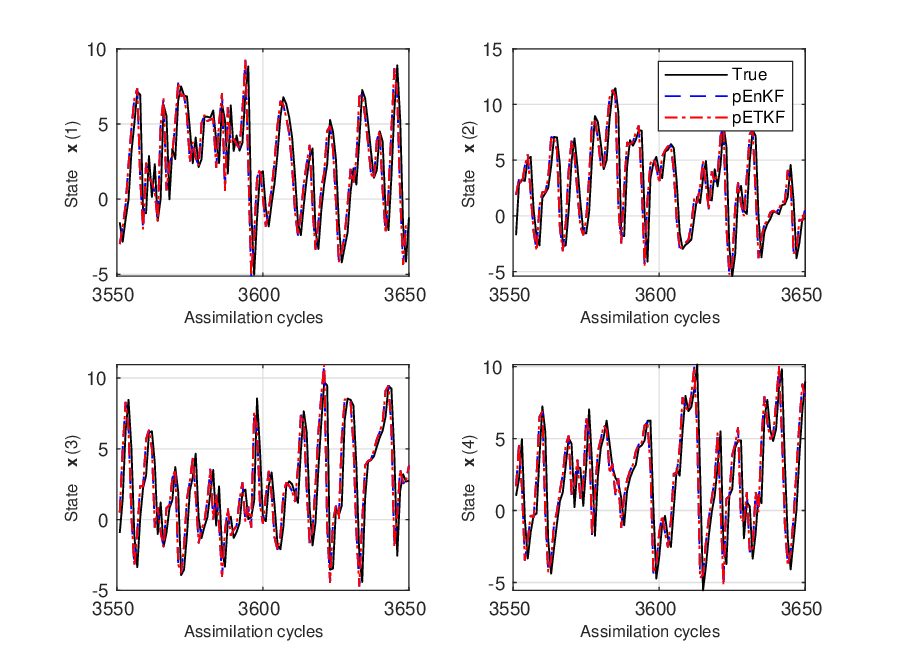}}  
\caption{Tracking of the first $4$ state variables with pSEnKF and pETKF within the last $100$ assimilation cycles.} 
 \label{figure_2}
\end{figure}

\begin{figure}[H] 
\centering 
\centerline{\includegraphics[width=10cm]{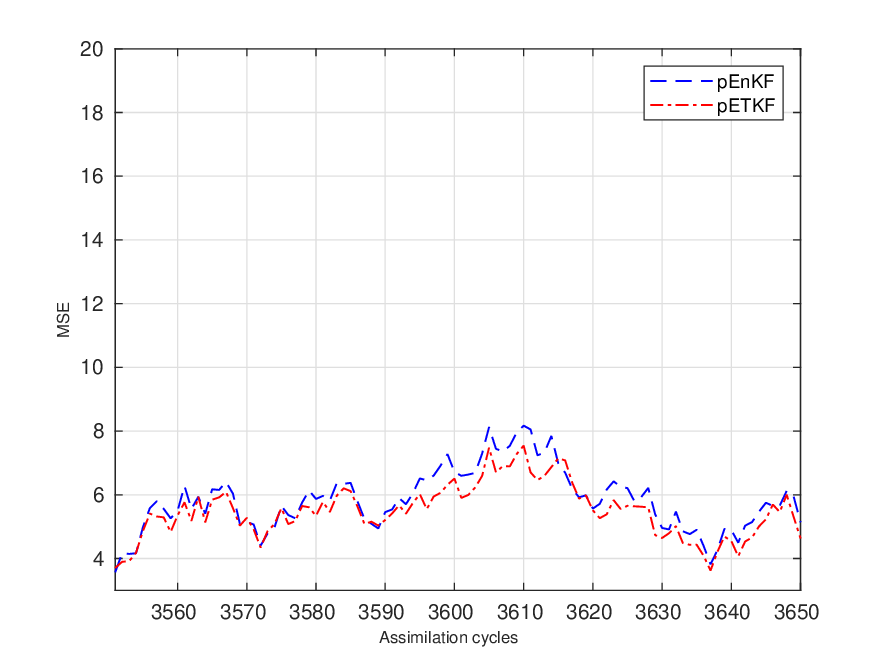}}  
\caption{Time-evolution of the MSE of the analysis state estimate as provided by pSEnKF and pETKF within the last $100$ assimilation cycles.}  
 \label{figure_3}
\end{figure}

Fig. \ref{figure_2} plots the first four components of the reference state during  the last $100$ assimilation cycles (for  better readability), and their analysis estimates as they result from the two proposed filters.  As can be seen, the reference states are well tracked by both algorithms. 
   The MSEs of the full state analysis estimates are further presented in Fig. \ref{figure_3}, suggesting a slightly better accuracy for the (deterministic) pETKF, as it supports better small ensembles (here $M=20$), compared to the (stochastic) pSEnKF.

 The second set of experiments is intended to asses the behavior of the proposed filters and to evaluate their performances against SEnKF and ETKF using empirically tuned inflation and localization, in two different  observational scenarios: full (i.e., all state variables are observed),  and spatially sparse,   assimilating  half   (i.e., odd-indexed  variables are observed) and quarter (i.e., every fourth  variable is observed) of the state variables. 
 SEnKF uses  a covariance localization based on the fifth-order correlation function given in \cite{GaspariC1999} \citep{hamill-et-al-2001-mwr}, and  ETKF uses the  local analysis approach \citep{paper-houtekamer-Mitchell-mwr-1998}.  
 In each scenario,   
  a series of sensitivity experiments is performed with different  levels  of  noise in the data (${\rm SNR} =10, 15$ dB,  respectively corresponding to $r = 1.87, 0.6$) and  different ensemble sizes ($M = 10, 20, 30$). 
   The  root MSE (RMSE) misfit between the  reference states and their analysis estimates, averaged over the whole assimilation  period, is used to evaluate the  performances of the filters. 
 
\subsection{Full observational scenario}  

\begin{figure}[H] 
\centering 
\centerline{\includegraphics[width=10cm]{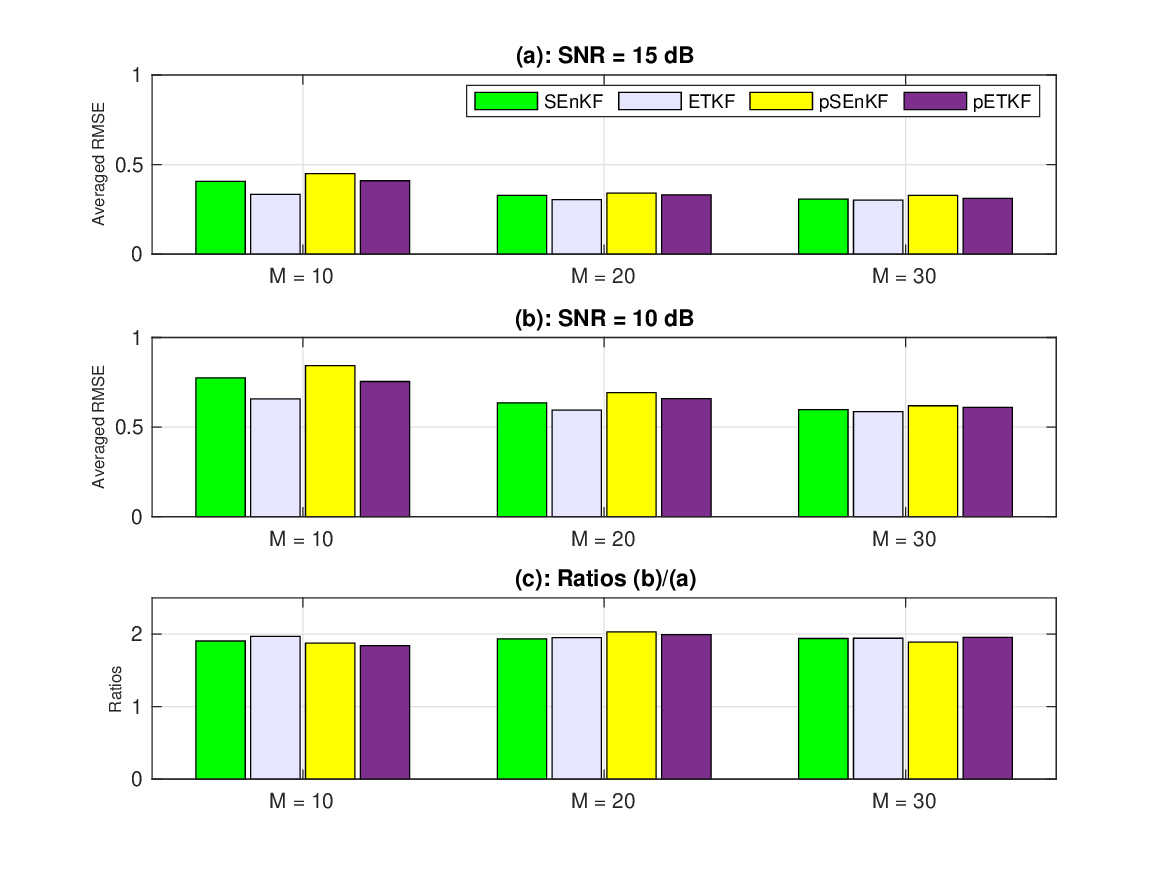}}
\caption{Averaged ${\rm RMSEs}$ of the state analysis estimates as provided by SEnKF, ETKF, pSEnKF and pETKF applied, with different ensemble sizes, on observations of ${\rm SNR} = 15$ dB (Panel (a)) and $10$ dB (Panel (b)). Panel (c) reports the  ratios of ${\rm RMSEs}$ in Panel (b)  w.r.t. those in Panel (a). 
} 
 \label{figure_4}
\end{figure}

The averaged analysis RMSEs  resulting from  the four filters when observations of all model variables are assimilated are presented  in Fig. \ref{figure_4}. Panels (a) and (b) represent the cases of ${\rm SNR} = 15$ dB and $10$ dB,   
  respectively, and Panel (c)   the corresponding ratios, i.e., the ratios of the RMSEs obtained for   ${\rm SNR} = 10$ dB  w.r.t. those associated to the case ${\rm SNR} = 15$ dB. The RMSEs of the pSEnKF and pETKF in Fig. \ref{figure_4} (ditto for Figs. \ref{figure_5}, \ref{figure_6} and \ref{figure_7} below) are obtained after trial-and-error tests using different values  of partitions' size, i.e.,  $d_p = 1, 2, 5$ for $M=10$, $d_p = 1, 2, 5, 10$ for $M=20$, and $d_p = 1, 2, 5, 10, 20$ for $M=30$.  The ``best'' values of $d_p$ here, i.e., those that correspond to the RMSEs  in Fig. \ref{figure_4}, are reported in Table  \ref{table_1}.

As can be seen from Table  \ref{table_1}, the  $d_p$ values that achieve the lowest RMSEs  increase with the ensemble size, $M$. Following the discussion in Section \ref{subsec-discussions},   this is due to the back-to-back approximation (VB density approximation, which improves with increased $d_p$, combined  with ensemble  approximation, which improves with increased $M$) on which these filters are based. This combination suggests that  obtaining  the lowest RMSE when  increasing $M$  is conditioned by an increase of $d_p$, subject to $d_p$ remaining smaller than $M$ (in order to avoid the need of  localization).  
 On the other hand,  Table  \ref{table_1} also suggests that, for a fixed $M$, the lowest RMSE 
 can be achieved when the state partition corresponds to either the largest $d_p$ value (among the tested ones) or to the one that precedes it. Taking for instance pSEnKF and ${\rm SNR} = 10$ dB: with $M=20$ members the lowest RMSE is achieved when  $d_p = 10$ (the largest value among the tested ones: $\{ 1, 2, 5, 10 \}$), and with $M=30$ it is achieved when $d_p = 10$ (the second largest value among those tested: $\{ 1,2,5,10,20 \}$). 
  Yet, it is important to note that increasing the ensemble can lead to better results for $d_p = 20$ compared to $10$. In this regard,  additional experiments with a  larger ensemble ($M=40$) have led to ${\rm RMSE} = 0.61628$ 
    for $d_p = 10$ and    $0.60297$ for $d_p = 20$.

The RMSEs reported in  Fig. \ref{figure_4} suggest that, as expected, all the filters behave better with large  ensembles  and  less noisy  observations.    The filters further overall   exhibit comparable performances,  providing  RMSEs that are more or less close to each other, as will be further discussed below.

As Panels (a) and (b) show, when a  small ensemble is used  (i.e., $M =10$ members),  the deterministic ETKF and pETKF filters    outperform their stochastic counterparts, regardless of the level of the noise  in the observations \citep{hoteit-et-al-2015}.  Such a difference becomes  less significant  with larger ensembles.   

As Panel (c) shows, all the filters exhibit  approximately twice  larger errors when passing from assimilated observations with ${\rm SNR} = 15$ dB  to  those with ${\rm SNR} = 10$ dB.

\begin{center}
 \begin{table*}[h!] \vskip .3cm  
 \centering 
\caption{The  values of $d_p$ that correspond to the errors plotted   in Fig. \ref{figure_4}.  
}
\vspace{1mm}
\begin{tabular}{lccc c ccc} \hline
  &  \multicolumn{3}{c}{${\rm SNR} = 15$ dB} &   &  \multicolumn{3}{c}{${\rm SNR} = 10$ dB} \\ 
 \cline{2-4} \cline{6-8} 
& $M = 10$ & $M=20$ & $M=30$ &  &   $M=10$ & $M=20$ & $M=30$     \\ \hline
{pSEnKF} & $5$ &  $10$  &  $20$  
&    & 
 $5$ &  $10$  
  & $10$   \\ 
{pETKF} & $5$ & $10$ &  $20$ &    & 
$5$ &   $10$  & $20$  \\ 
	\hline
	\end{tabular} 
\label{table_1} 	
\end{table*}
\end{center}

\subsection{Spatially sparse observational scenario}  

\begin{figure}[H] 
\centering 
\centerline{\includegraphics[width=10cm]{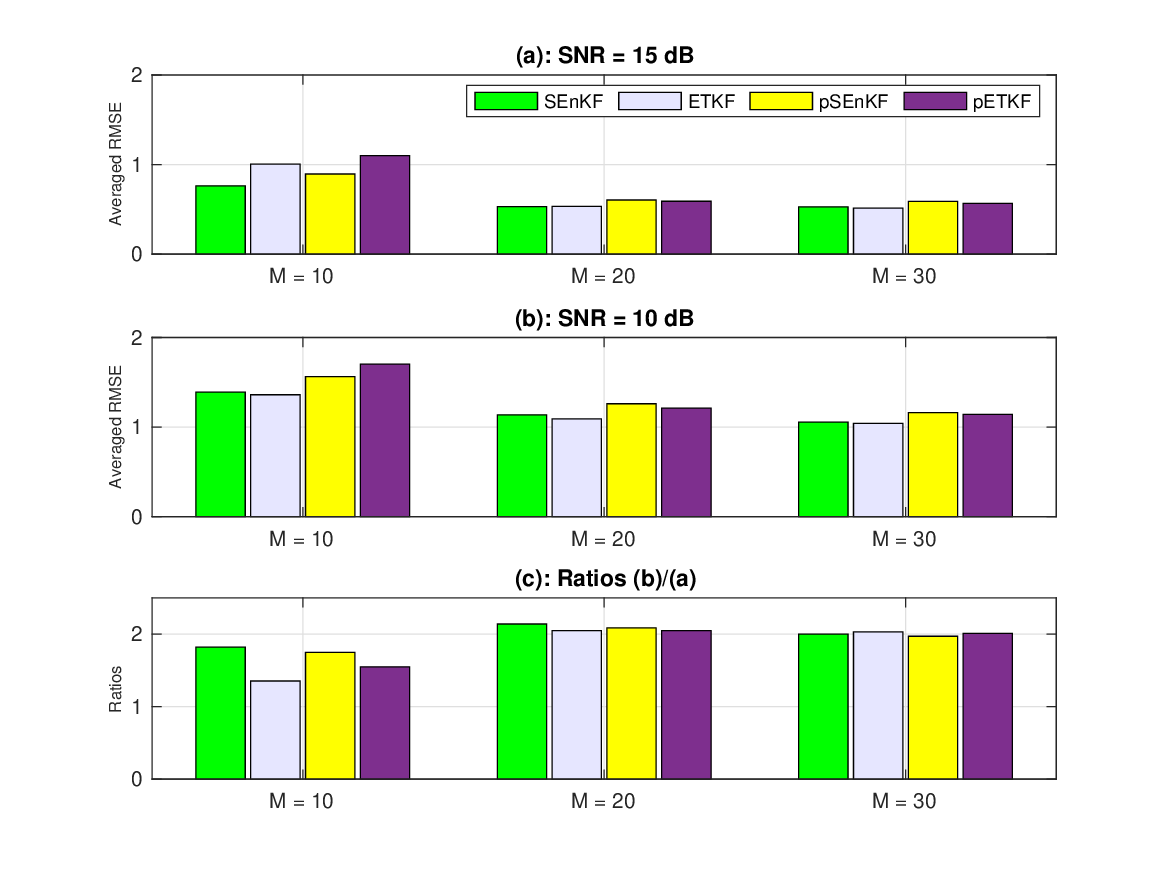}}  
\caption{As in Fig. \ref{figure_4}, but with observing  odd-indexed state variables  only.} 
 \label{figure_5}
\end{figure}

We start by the scenario where half of the observations are assimilated and plot the resulting averaged analysis REMSs in Fig. \ref{figure_5}.  
 The partitions' sizes, $d_p$,  that correspond to these (minimum) errors following  trial-and-error tests  are  reported in Table  \ref{table_2}. 
 
 The results in Table  \ref{table_2} are   close to those in Table  \ref{table_1}. Notably,  the values of $d_p$ that achieve the lowest RMSEs  increase with the ensemble size, $M$, and   for each   $M$, the state  partitioning  that provides the lowest  RMSE  corresponds to one of the two largest values of  $d_p$ among the tested ones.

The RMSEs in Fig. \ref{figure_5} are found to be larger than those in Fig. \ref{figure_4}, which is expected as less observations are assimilated in this experiment.     They further generally decrease with the ensemble size and  less noisy  observations.

 Fig. \ref{figure_5} also shows that  the four filters overall   exhibit comparable performances,  providing  RMSEs that are more or less close to each other.  Specifically, when $M =10$ members are used, the proposed pSEnKF and pETKF  slightly underperform their standard counterparts (most notably when ${\rm SNR = 10}$ dB). 
 Such a behavior becomes  less significant  with larger ensembles ($M =20$ and $30$ members).

\begin{figure}[H] 
\centering   
\centerline{\includegraphics[width=10cm]{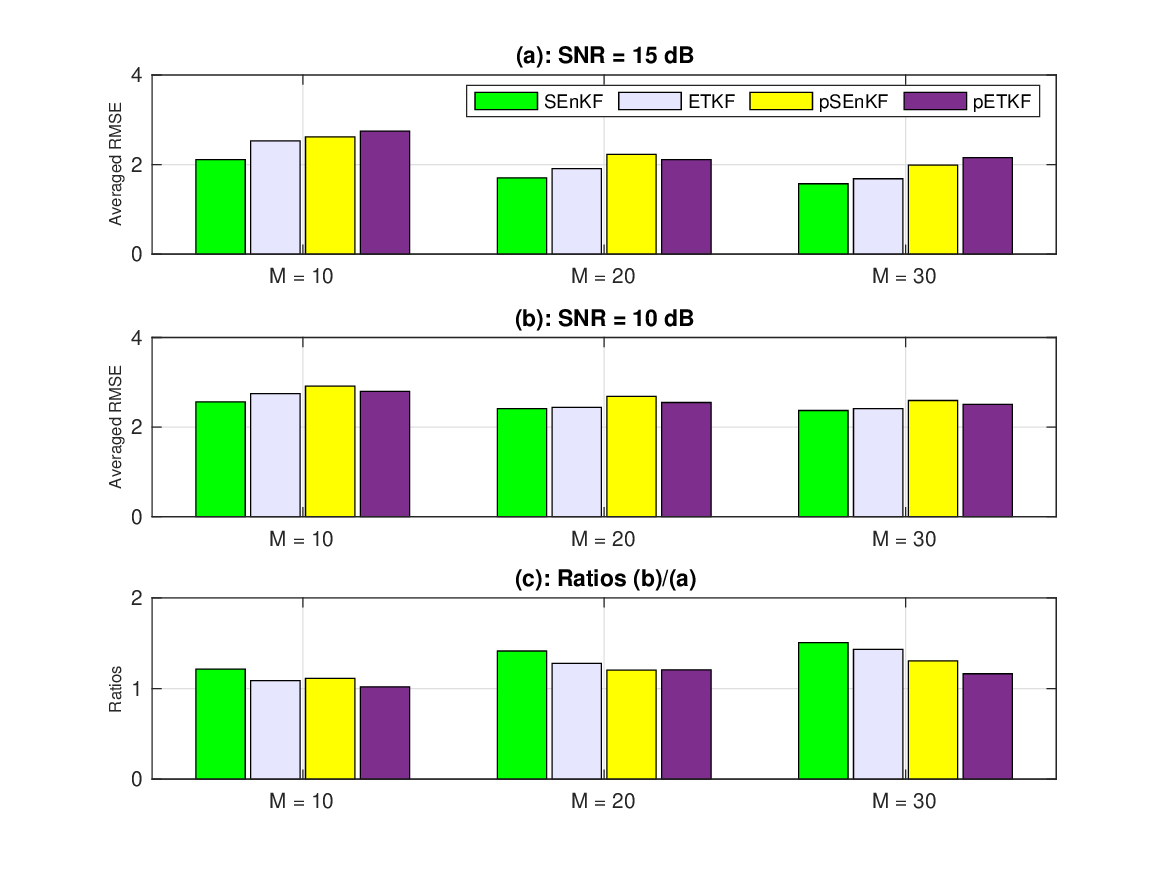}}  
\caption{As in Fig. \ref{figure_4}, but with observing  every fourth state variable.} 
 \label{figure_6}
\end{figure}

  Overall, similar results are obtained  in the quarter observational scenario, as can be noticed from  Fig.   \ref{figure_6} and Table  \ref{table_3}, yet with the  exception of the RMSEs being  less sensitive to more noise in the observations.

\begin{center}
 \begin{table*}[h!] \vskip .3cm  
 \centering 
\caption{The  values of $d_p$ that correspond to the errors plotted   in Fig. \ref{figure_5}.  
}
\vspace{1mm}
\begin{tabular}{lccc c ccc} \hline
  &  \multicolumn{3}{c}{${\rm SNR} = 15$ dB} &   &  \multicolumn{3}{c}{${\rm SNR} = 10$ dB} \\ 
 \cline{2-4} \cline{6-8} 
& $M = 10$ & $M=20$ & $M=30$ &  &   $M=10$ & $M=20$ & $M=30$     \\ \hline
{pSEnKF} & $5$ & $10$ 
& $10$ &    & 
 $5$ &  $10$  
   & $10$   \\ 
{pETKF} &    $2$ & $5$ 
&  $20$ &    & 
$2$ &   $10$  & $20$  \\ 
	\hline
	\end{tabular}
 \label{table_2}	
\end{table*}
\end{center}

\begin{center}
 \begin{table*}[h!] \vskip .3cm  
 \centering 
\caption{The  values of $d_p$ that correspond to the errors plotted   in Fig. \ref{figure_6}. 
}
\vspace{1mm}
\begin{tabular}{lccc c ccc} \hline
  &  \multicolumn{3}{c}{${\rm SNR} = 15$ dB} &   &  \multicolumn{3}{c}{${\rm SNR} = 10$ dB} \\ 
 \cline{2-4} \cline{6-8} 
& $M = 10$ & $M=20$ & $M=30$ &  &   $M=10$ & $M=20$ & $M=30$     \\ \hline
{pSEnKF} & $5$ & $10$ & $10$ &    & 
 $5$ & 
   $5$   & $10$   \\ 
{pETKF} & $2$ & $5$ &  $10$ &    & 
$5$ &  
  $5$  & $10$  \\ 
	\hline
	\end{tabular}
 \label{table_3} 	
\end{table*}
\end{center}

  We further assess the statistical consistency of the proposed partitioned approach by comparing the analysis ensemble spread against  the (time-series) analysis RMSE. For that, we use pSEnKF with  $d_p = 10$ (i.e., four partitions) and $M=30$ in the configuration ``SNR = $15$ dB, half observational scenario'' (Panel (a) in Fig.  \ref{figure_5}). 
  The goal is to determine whether the partition-wise update results in under- or over-dispersive ensembles and to identify potential artifacts/imbalances in uncertainty  near the partition boundaries. 
 To achieve this, we categorize the state indices/elements into four subsets based on their observational status (observed vs. unobserved) and their proximity to partition boundaries. Components are considered as ``Close'' if they lie within two grid points of a boundary, while the remaining (interior) components are classified as ``Far''.  
 
 Such four subsets correspond to: $\{$Observed and Close$\}$, $\{$Unobserved and Close$\}$, $\{$Observed and Far$\}$, and $\{$Unobserved and Far$\}$.     For each assimilation cycle, the analysis ensemble spread and RMSE are computed independently using pSEnKF across these  subsets. 

\begin{figure}[H] 
\centering   
\centerline{\includegraphics[width=10cm]{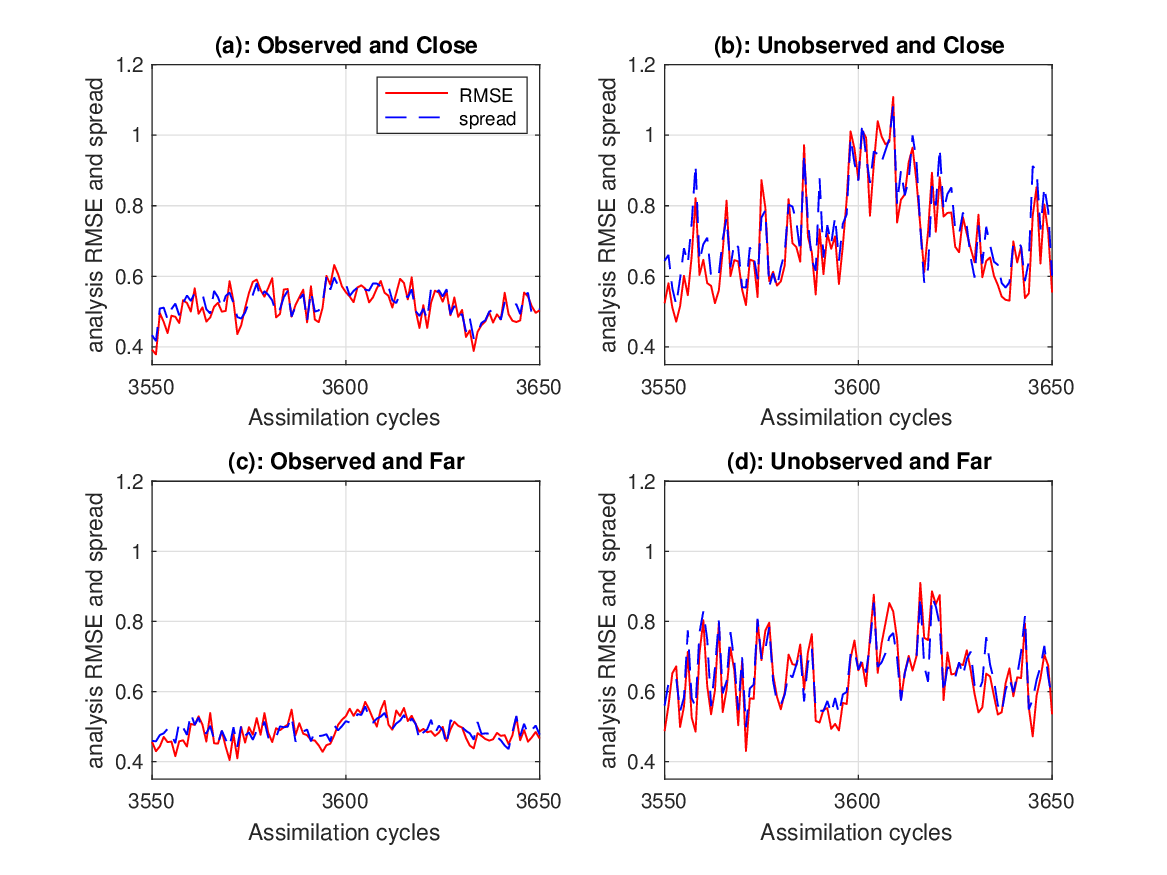}}  
\caption{Time-evolution of the analysis RMSE  and ensemble spread as provided, within the last $100$ assimilation cycles, by pSEnKF implemented with $M=30$ members and $d_p = 10$ in the half observational scenario with SNR $= 15$ dB (one scenario of Panel (a) in Fig. \ref{figure_5}).  The results are averaged independently over four subsets of state variables defined based on their observability and proximity to partition boundaries. 
} 
 \label{figure_6_1}
\end{figure}

The results are reported in Fig. \ref{figure_6_1}. As expected, unobserved variables (Panels b and d) exhibit larger errors and spread compared to observed ones (Panels a and c). The ensemble spread further demonstrates a clear consistency with the RMSE as it closely tracks it in all subsets. In particular, the filter does not suffer from under-dispersion at the partition boundaries (Panels a and b), even for unobserved variables. In other terms, the partitioning  did not induce noticeable boundary artifacts in the spread and error. This suggests that the iterative adjustment in the analysis step, combined with the mixing provided by the model dynamics in the forecast step,  effectively propagates information across partitions and  preserves a coherent ensemble spread globally, avoiding artificial discontinuities at the boundaries of the partitions. 

\subsection{More challenging scenarios} 
  
The  case of ${\rm SNR} =10$ dB, $M=10$ and $\delta_o = 4 \delta_m$ in the quarter observational scenario (i.e., Fig. \ref{figure_6}) is the most  challenging  among those presented so far    
(this case will hereafter refer to Case 0). 
 Here, we examine the robustness of the filters to even more challenging case scenarios favouring additional systematic errors, through increasing  the observations' temporal sparsity and using wrong models in the filtering algorithms. These include four case scenarios,  sorted below according to their complexity:  
\begin{itemize}
\item {Case 1:}  Case 0,  with $\delta_o = 8 \delta_m$ instead of $4 \delta_m$;  
\item {Case 2:}  Case 1,    with a bias of $0.3F$ induced in the state model during the filtering;  
\item {Case 3:} Case 2, with a  bias of $0.3 {\bf H}{\bf x}_0$ induced in the observation model during the filtering;      
\item {Case 4:}  Case 3, with reducing the (true) observation noise covariance  ${\bf R}$ by $30\%$ during the filtering  (i.e.,   $0.7 {\bf R}$ is used as the prescribed observation noise  covariance in assimilation, while ${\bf R}$ is the true observation noise covariance used in generating observations from truth).     
\end{itemize}  

\begin{figure}[H] 
\centering 
\centerline{\includegraphics[width=10cm]{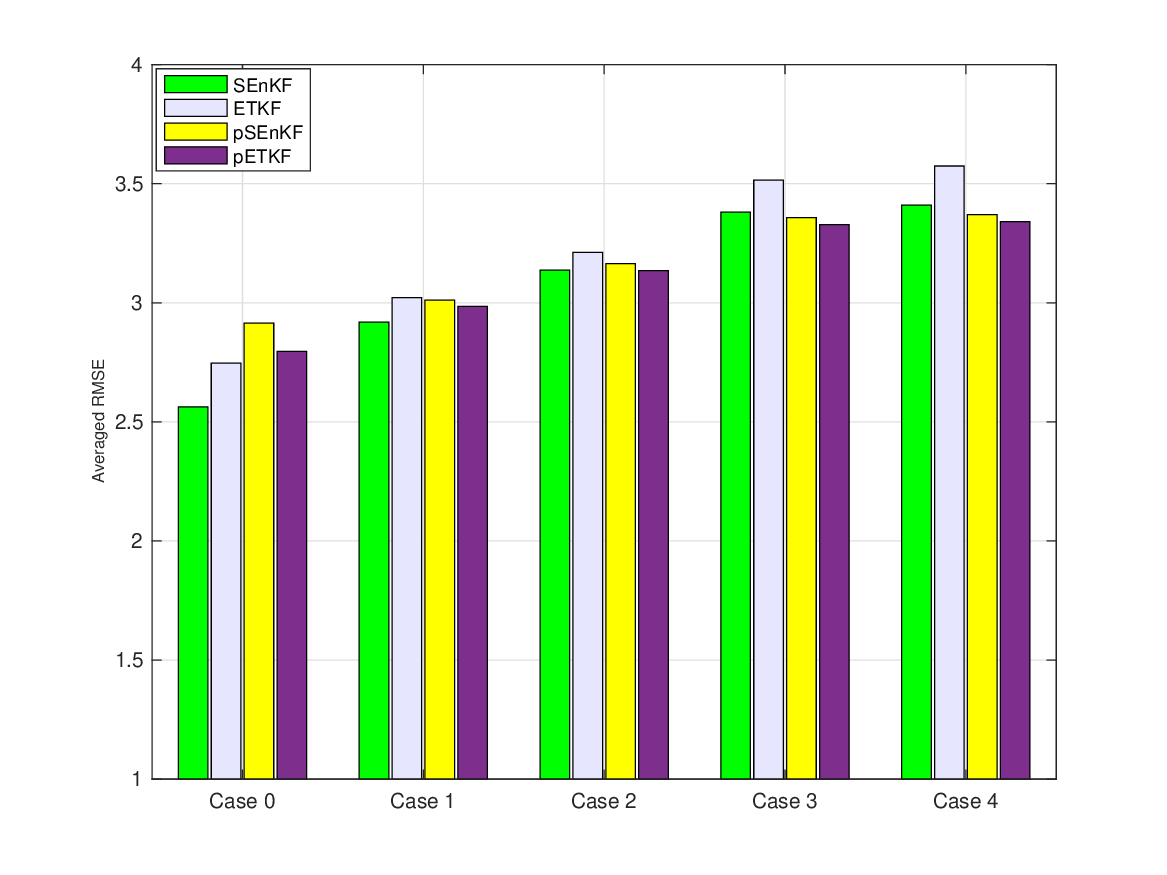}}  
\caption{Averaged ${\rm RMSEs}$ of the state analysis estimates as provided by SEnKF, ETKF, pSEnKF and pETKF in different challenging case scenarios. } 
 \label{figure_7}
\end{figure}

Fig. \ref{figure_7} plots  the resulted averaged analysis RMSEs. As can be seen from comparing Case 1 with Case 0, all the  filters are  clearly sensitive to less frequent observations.   
   In Case 1, the proposed filters  still underperform the standard SEnKF,  in a similar way to Case 0   although less significantly. Compared to ETKF, the pSEnKF is however found to achieve a very closer behavior and the pETKF became even slightly more accurate.  

Comparing Case 2 with Case 1, it turns out that  the effect of a bias in the state model  is slightly  more pronounced on the standard filters than on the proposed ones. This may reflect   a slightly more  deterioration of  the forecast background in the standard filters, caused by the use of the erroneous model. 
  One further notes that in Case 2, the pSEnKF and pETKF behaviors became  closer to that  of SEnKF,  
 and clearly better than that of ETKF. 

 All the filters are found to be significantly affected where a bias is induced in the observation model too (Case 3), exhibiting RMSEs that significantly increase when moving from Case 2 to Case 3. 
  In Case 3, the two proposed filters are further found to  significantly outperform the ETKF and  slightly the SEnKF.

As can be seen from Cases 3 and 4, the proposed pSEnKF and pETKF filters, and to a slightly lesser extent the standard filters,  seem to be  robust to increased confidence in wrong observations  
 (the effect of  deflating  ${\bf R}$ by $30\%$ is more or less insignificant).   
 Furthermore,  the gap between the RMSEs of the standard filters and those of the proposed ones  has been found to increase when moving from Case 3 to Case 4. 
  The outperformance of the proposed filters in such challenging scenarios likely stems from
their design: they use the same observational information as the standard filters but
apply it to smaller state partitions, updated iteratively. This localized and iterative
update scheme reduces the impact of parameter errors and enhances stability under challenging conditions.

 It is further worth assessing the relevance of the assimilation through comparisons  against baselines, such as the climatological level of variability, $\sigma_{\rm clim}$, which gives an idea on how large the natural fluctuations of the truth (nature run) are, and the time-series RMSE of the observations, ${\rm rmse}_{\rm obs}$. These  are respectively defined as, 
\begin{eqnarray*} 
\sigma_{\rm clim} & = & \frac{1}{\sqrt{d_x}}  \| \boldsymbol{\delta}\|,  \\ 
{\rm rmse}_{\rm obs}(n) & = & \frac{1}{\sqrt{T}} \| {\bf y}_n - {\bf H}_n {\bf x}_n \|, 
\end{eqnarray*} 
where the $i$-th element  of the $(d_x \times 1)$ sized vector  $ \boldsymbol{\delta} $ is defined as 
$\delta^i = \sqrt{ \frac{1}{T} \sum_{n=0}^{T-1} (x_n^i - \overline{x}^i)^2}$,   ${\bf x}_n$ and ${\bf y}_n$ respectively denote  the reference state and the observation at assimilation time $n$, and $\overline{x}^i$ the (temporal) average of ${\{ x_n^i \}}_n$. 

\begin{figure}[H] 
\centering  
\centerline{\includegraphics[width=10cm]{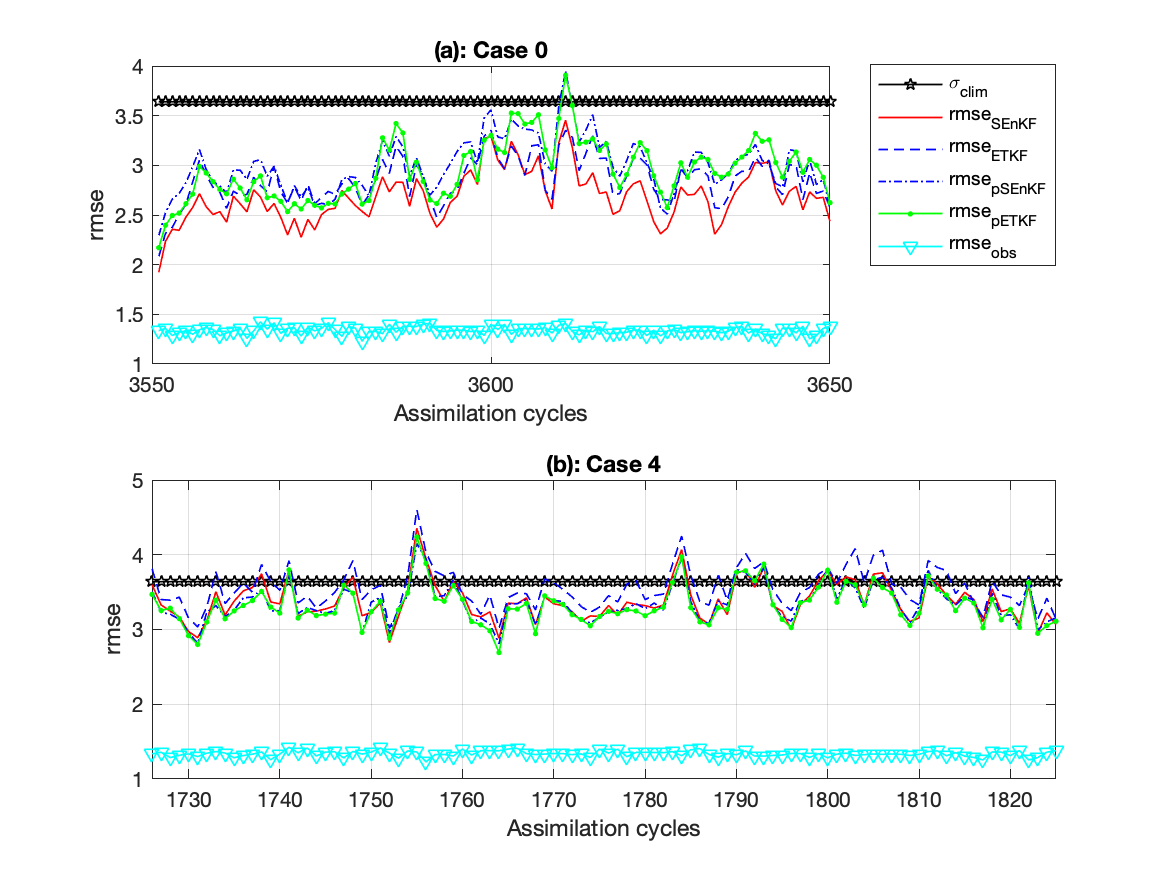}}  
\caption{The climatological level of variability and the time-series  ${\rm rmse}$ of the observations and the state analysis estimates  provided by the four filters in Cases 0 and 4, within the last $100$ assimilation cycles. } 
 \label{figure_7_1}
\end{figure}

Fig. \ref{figure_7_1} plots these quantities  and the temporal ${\rm rmse}$ errors of the state analysis estimates provided by the four filters in Cases $0$ and $4$,   within the last $100$ assimilation cycles. According to the results in Panel (a), although the filters, including the standard ones, seem to not fit the data very well in such a challenging scenario (as their analysis ${\rm rmse}$ errors exceed ${\rm rmse}_{\rm obs}$),  they remain clearly useful relatively to   the climatological mean (as their analysis ${\rm rmse}$ errors  are clearly smaller than  $\sigma_{\rm clim}$).  In Panel (a), one also recognizes  the order of the analysis  errors indicated in Fig. \ref{figure_7}, i.e., overall SEnKF outperforms  ETKF, then  pETKF and    pSEnKF.

  When the filters further run with larger assimilation cycles and perturbed parameters (Panel (b)), the  analysis errors increase, compared to those in Panel (a), but overall remain below $\sigma_{\rm clim}$, suggesting that the assimilation results remain meaningful even in a such challenging scenario.

 \subsection{A case of non-uniform state partitioning} 
    As previously outlined, the proposed schemes are  straightforward to adapt to the case where the $K$ partitions ${\bf x}_n^k$  have different sizes $d_{p^k}$, by only replacing in Algorithms 1 and 2 $d_p$ by $d_{p^k}$ and $K d_p$ by $\sum_{k=1}^K d_{p^k}$.  Two configurations are considered here: 
\begin{itemize}
\item {\sl Configuration 1:} The state is split into $K=3$ partitions of dimensions  $[d_{p^1}, d_{p^2}, d_{p^3}]$ that can be equal to  ${\cal D}_3^1 = [15, 13, 12]$,   ${\cal D}_3^2 = [17, 11, 12]$, ${\cal D}_3^3 = [16, 15, 9]$ or ${\cal D}_3^4 = [10, 20, 10]$. 

\item {\sl Configuration 2:} The state is split into $K=4$ partitions of dimensions  $[d_{p^1}, d_{p^2}, d_{p^3}, d_{p^4}]$ that can be equal to  ${\cal D}_4^1 = [9, 2, 15, 14]$,   ${\cal D}_4^2 = [6, 15, 9, 10]$, ${\cal D}_4^3 = [10, 10, 10, 10]$ or ${\cal D}_4^4 = [10, 8, 13, 9]$. 
\end{itemize}     
   For each case, the  proposed filters are run using $M=30$ members in a non-uniform half observational scenario, in which the variables ${\{y_n^i; i =  1, 4, 5, 8, 10, 12, 13, 14, 17, 20, 24, 25, 27, 28,}$ ${ 31, 32, 34, 35, 37, 40 \}}$ are observed, ${\rm SNR} =15$ dB     and $\delta_o = 4 \delta_m$. The  standard filters are also run  
   with ``optimally'' tuned inflation and localization.

\begin{figure}[H] 
\centering  
\centerline{\includegraphics[width=10cm]{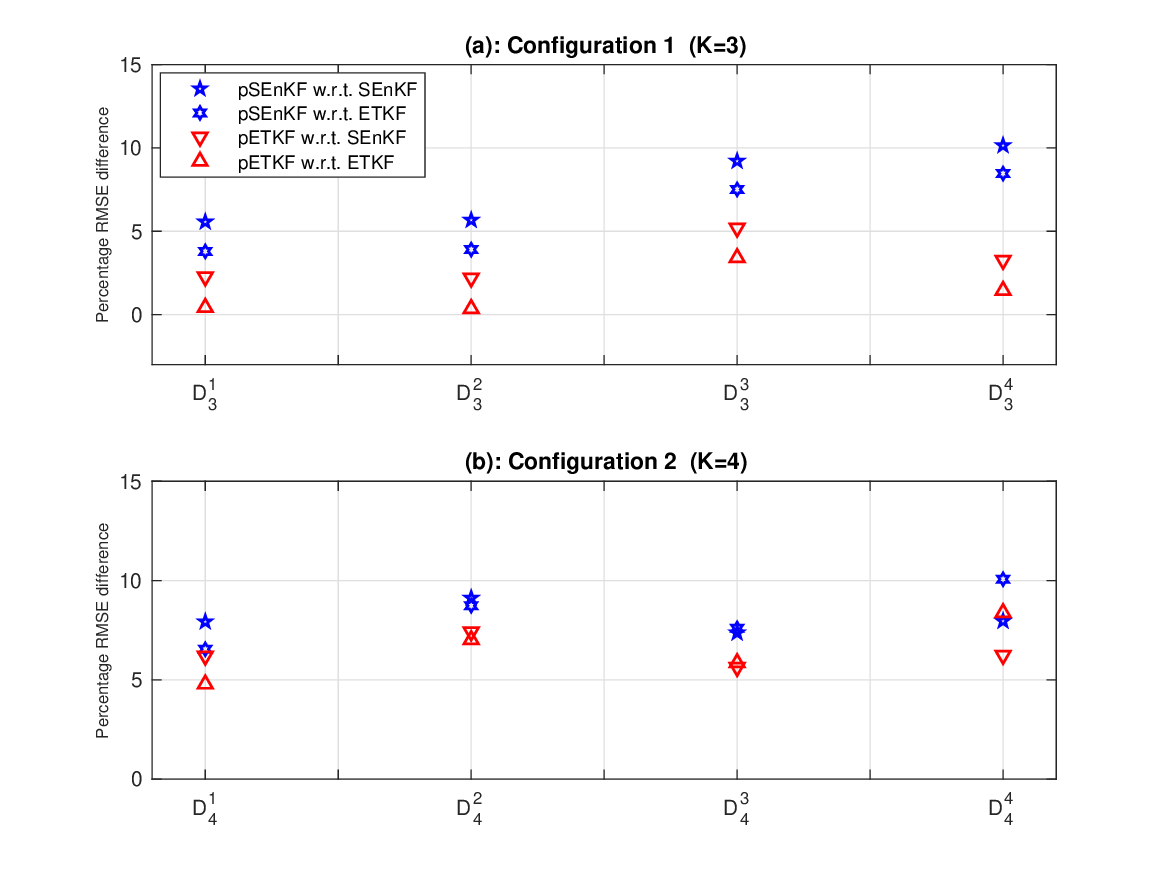}} 
\caption{The difference (in $\%$) between the averaged state analysis RMSEs  provided by the  standard SEnKF and ETKF  and those provided by the proposed pSEnKF and pETKF splitting the state non-uniformly following Configurations 1 (Panel (a)) and 2 (Panel (b)), all using $M=30$ members in a non-uniform half observational scenario with ${\rm SNR} = 15$ dB and $\delta_o = 4 \delta_m$.  
} 
 \label{figure_8}
\end{figure}

   Fig.  \ref{figure_8} plots the percentage of the difference between the averaged state analysis RMSEs as resulting from  the proposed filters and those obtained by the standard filters, defined as, 
    \begin{equation}
   \nonumber 
\frac{\rm RMSE_{\rm proposed \, filter}  - RMSE_{standard \, filter}}{\rm RMSE_{\rm standard \, filter}} \times 100. 
   \end{equation}      
    Overall, the  proposed filters in both configurations provide close performances to the  standard ones. 
    
      As can be seen from Panel (a),  with  all three-part state partitions, the pETKF achieved RMSEs  that do not exceed $5.2 \%$ of the RMSEs  provided by  both standard filters. 
   It even behaved as good as  ETKF when it is implemented with  ${\cal D}_3^1$ or ${\cal D}_3^2$. The pSEnKF  performs less than  pETKF, achieving its best performance when implemented with ${\cal D}_3^1$ or ${\cal D}_3^2$, cases in which it has suggested RMSEs that are respectively  $3.79 \%$ and $3.9 \%$ larger than that  of ETKF. 
   
   In the  four-part partitioning  configuration (Panel (b)),  the percentage errors do not exceed $10 \%$, like in Panel (a), but are overall larger than those in Panel (a). 
   This suggests that in this observational scenario, it is generally more beneficial to split the state into three partitions instead of four.

 \subsection{Illustration through assimilating a single observation} 
  
\begin{figure}[H] 
\centering 
\includegraphics[width=10cm]{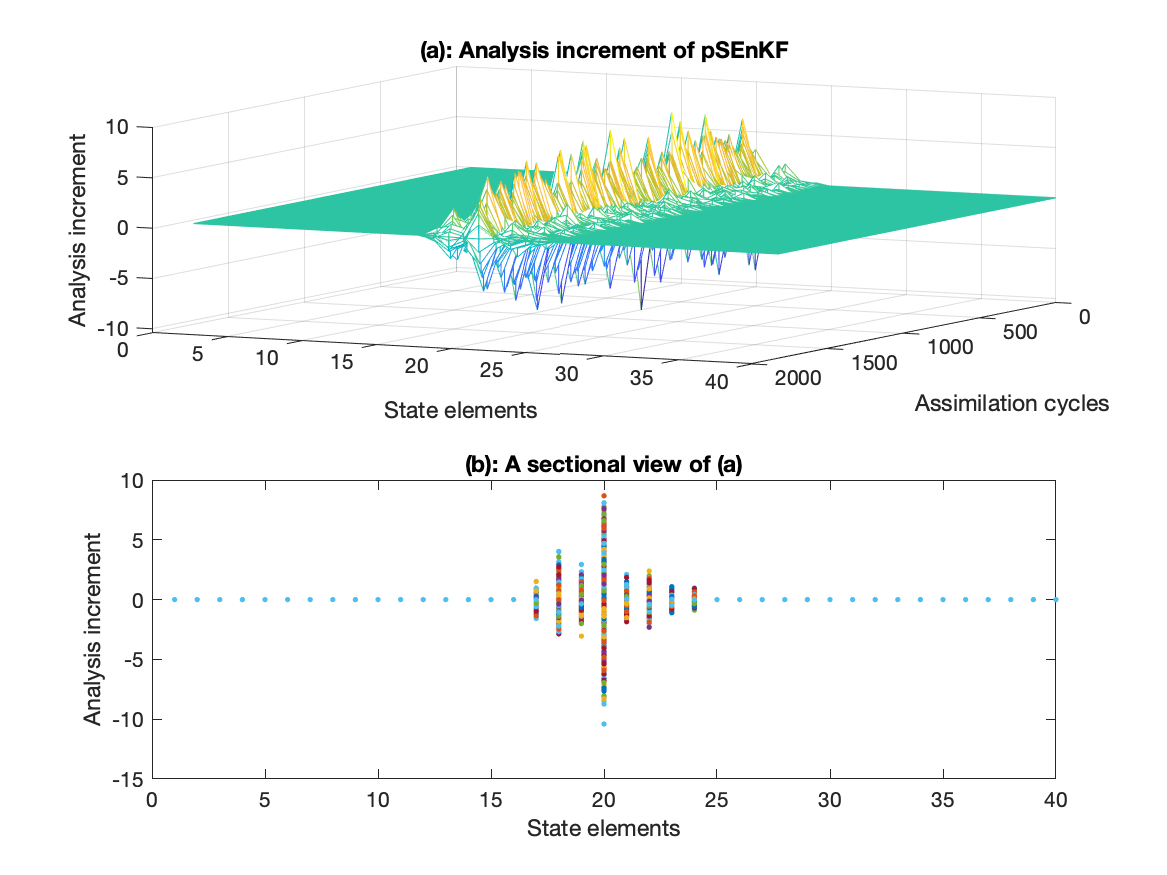}
\caption{a) Time-evolution of the analysis increment of all state elements as provided, within the last 100 assimilation cycles, by pSEnKF implemented with $M=30$ members, $d_p = 8$ and ``optimal'' inflation (with factor $\alpha = 1.1$). 
 (b) is a sectional view of (a). Only the $20$th state element is observed and assimilated.  } 
 \label{figure_9}
\end{figure}

  To illustrate how  the proposed approach deals with  (sampling and systematic) noises,  single observation experiments are conducted,   observing only  the $20${\sl th} state variable with ${\rm SNR} = 15$ dB. Fig. \ref{figure_9}  plots the time evolution of the  analysis increment (the analysis mean mines the forecast mean) over the last $100$ assimilation cycles as provided by pSEnKF using $M = 30$ members,   $\delta_o = 4 \delta_m$ and splitting the state vector into partitions of size $d_p = 8$ (pETKF have led to similar results, but they are not included here  to limit the length  of the paper). As can be seen from Fig. \ref{figure_9}, the analysis increments are localized around the position of the observation, meaning that the localization works as expected. 
 
  Fig. \ref{figure_9} also shows that during the analysis cycle, the information from the observation has been retained within the state partition in which it is located (i.e., the partition ${\bf x}_n^3 = (x_n^{17}, \cdots, x_{n}^{24})$), and did not propagate across the other partitions. This is due to i)  the removal of the forecast cross-covariances (i.e., to ${\bf G}_n^k = {\bf 0}$), combined with ii)   the particular structure of ${\bf H}$ ($ = [0, \cdots, 0, 1, 0, \cdots, 0]$) which arises in this single-observation case. 
   Indeed, since ${\bf G}_n^k = {\bf 0}$, the update of the state partitions is performed by the analysis Eq. (\ref{vb-trans-pdf-xk-mean-analysis-approx}) instead of Eq. (\ref{vb-trans-pdf-xk-mean-analysis}), and due to  the sparse structure of ${\bf H}$,  the Kalman gains ${\bf L}_n^k$ in Eq. (\ref{vb-trans-pdf-xk-mean-analysis-approx}) that are associated to all partitions, except  the one that contains the observation (i.e, ${\bf x}_n^3$), become null.  This results in analysis estimates that are identical to the forecast estimates (i.e., with no analysis increments) in the partitions that do not include the  observation.

\begin{figure}[H] 
\centering 
\includegraphics[width=10cm]{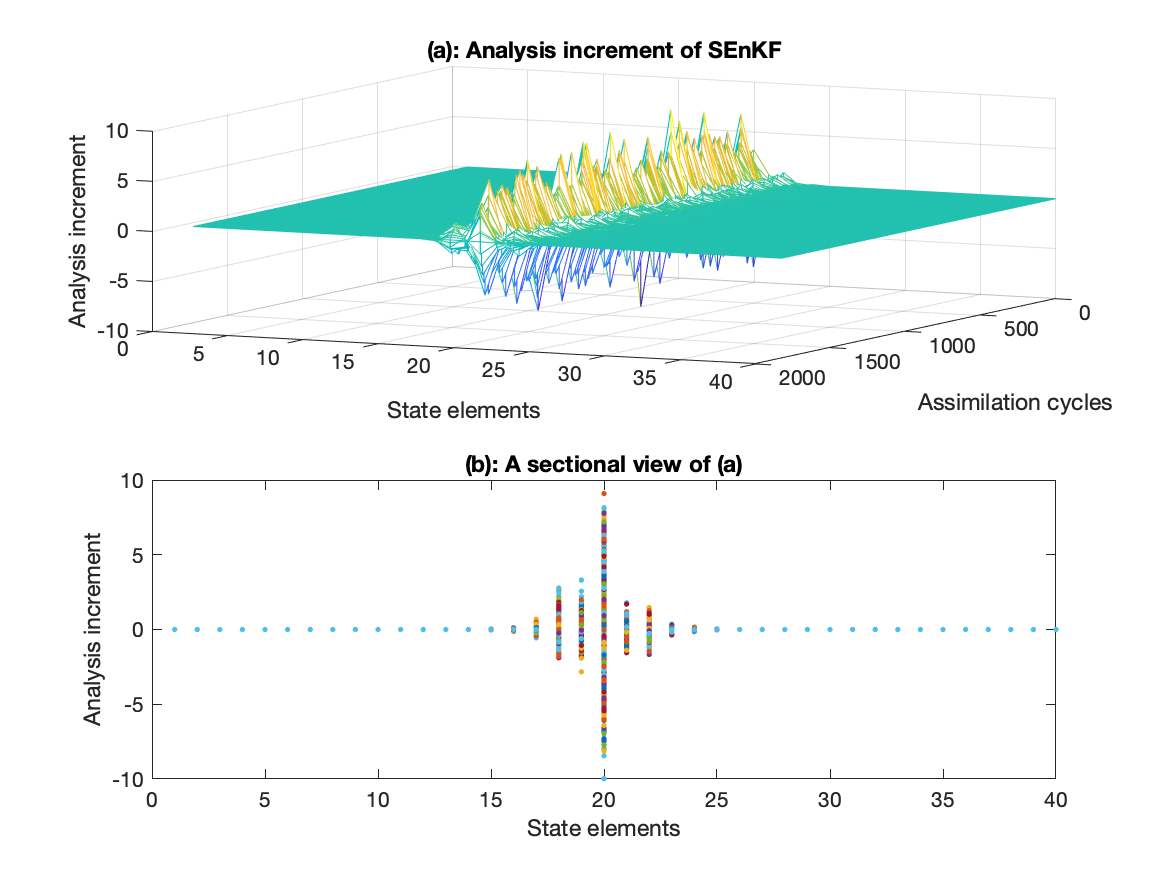}  
\caption{a) Time-evolution of the analysis increment of all state elements as provided, within the last 100 assimilation cycles, by SEnKF implemented with $M=30$ members, ``optimal'' inflation (with factor $\alpha = 1.1$) and localization (with length-scale $\ell_s = 2$). (b) is a sectional view of (a). Only the $20$th state element is observed and assimilated.} 
 \label{figure_10}
\end{figure}

 For a similar reason, the standard SEnKF with   tuned covariance localization (with an ``optimal'' length-scale  $\ell_s  = 2$) has been found to exhibit a similar behavior (cf. Fig.  \ref{figure_10}). More precisely, the relatively strong  localization ($\ell_s = 2$)   drastically limits the forecast cross-covariances, and    
   this, together  with the particular structure of ${\bf H}$, limits the propagation of the information in the observation only within the  close neighborhood.  In that respect, these two filters lead to comparable RMSEs: $3.6315$  
    for pSEnKF and $3.6245$     for SEnKF.

 Finally, to investigate  the effect of larger partition size in this single-observation scenario, we  repeated the  experiment of Fig. \ref{figure_9}  
 with $d_p = 20$ instead of $8$. Results not shown here indicate that the increment remains confined to the partition containing the observed component (no cross-partition long-range effects), while the within-partition spatial extent naturally broadens. RMSE slightly degrades  compared with the  $d_p = 8$ case (Fig. \ref{figure_9}),  
  but improves when $M$  is increased (RMSE $=3.6383$ and $3.6287$  respectively for $M = 40$ and $45$), consistent with sampling limitations when $d_p$ becomes large relative to $M$. 

 Overall, this confirms that larger partitions are most beneficial when supported by a sufficiently large ensemble, and suggests that $d_p$  effectively controls a discrete (partition-wise) localization scale, 
 subject to the usual ensemble-size sampling constraint: reliable within-partition covariance estimation requires $M$  to be sufficiently large relative to $d_p$ (in practice  $M > d_p$, and ideally $M \gg  d_p$).

\section{Conclusion} 
\label{ref-sec-6} 

This work proposes  an ensemble Kalman filtering (EnKF) approach that is inherently localized,  avoiding the need for   any auxiliary  localization technique. Its principle is based on first localizing the continuous analysis probability density function (pdf), then applying Gaussian-like ensemble Kalman  sampling procedures on the localized pdf. 
  The localization of the analysis pdf  consists of approximating it   by a product of independent marginal pdfs corresponding to small partitions of the state vector, using the variational Bayesian optimization. These marginals are then sampled following stochastic EnKF (SEnKF) and deterministic ensemble transform Kalman filtering (ETKF) procedures, but any other sampling procedure could be applied too. The rational behind the proposed approach relies  on the assumption that  selecting partitions with dimensions smaller than the ensemble size should boost performances without incorporating any external localization scheme, while mitigating for the effect of undersampling and systematic errors.

 Algorithmically speaking, the   filters  involve the same forecast step as the standard SEnKF and ETKF,  but different  analysis steps which consist in  iteratively  adjusting the standard   SEnKF   and ETKF updates of each partition,  with a shift by a linear combination of the  ensemble means of the other partitions. 
Numerical experiments are conducted with the Lorenz-96 model under various challenging settings and scenarios, demonstrating the relevance of the proposed filters in geophysical applications, being generally comparable to the SEnKF and ETKF with already tuned localization (and sometimes even better than them), both in terms of computational burden and estimation accuracy. 

 Future work will consider enhancing the new filters' performances by integrating computationally   reasonable approximations of the gain terms in  Eqs. (\ref{vb-trans-pdf-xk-mean})-(\ref{vb-trans-pdf-xk-cov}), and by mitigating  for the imposed posterior independencies between the state partitions following e.g. the approach of  \citep{Ranganath-et-al-2016}.  
 We will also consider integrating the online estimation of the inflation factor as e.g. in \citep{inflationPaper-2020}, with the goal to derive more autonomous ensemble filtering schemes, avoiding the need for trial-and-error tuning of this factor.  
 Other important directions for future work would be to apply the proposed inherently localized  approach on other deterministic ensemble filters, e.g., SEIK \citep{hoteit-et-al-2002},  LETKF \citep{hunt-et-al-physica-2007}, and to apply  all the new filters to  realistic oceanic and atmospheric data assimilation problems.


\section*{Appendix}
\label{AppendixA}

\vspace{-.01cm}  
\begin{center}
\textbf{The analysis step of the generic localized filter} 
\end{center} 
 
We show how (\ref{eq-trans-y-N-v0-vb})-(\ref{vb-trans-pdf-xk-cov-analysis})  can be derived  from (\ref{eq-vbIter-classic-partition}).  

The Gaussian  expression (\ref{eq-trans-y-N-v0}) of $p({\bf y}_n|{\bf x}_n)$ entails, 
\begin{equation} 
\label{Argument-pf-lik} 
p({\bf y}_n|{\bf x}_n)  \propto  \exp \left( -\frac{1}{2} \| {\bf y}_n - {\bf H}_n^{k-} {\bf x}_n^{k-} - {\bf H}_n^k {\bf x}_n^k\|_{{\bf R}_n^{-1}}^{2} \right). 
\end{equation}  
Applying the $\ln$ function to (\ref{Argument-pf-lik}), followed by the expectation w.r.t.  $\pi_n^{(i,i-1)}({\bf x}_{n}^{k^-}) = \prod_{j=1}^{k-1}  \pi_n^{i}({\bf x}_n^{j}) \times \prod_{j=k+1}^{K}   \pi_n^{i-1}({\bf x}_n^{j})$, then the $\exp$ function, one obtains, 
\begin{eqnarray}  
\nonumber 
{\cal L}_{{\bf y}_n}^i({\bf x}_{n}^k)  & \propto &    \exp   \left(\mathbb{E}_{\pi_n^{(i,i-1)}({\bf x}_{n}^{k^-})} [\ln p({\bf y}_n | {\bf x}_{n})] \right),   \\
\nonumber 
 & \propto & \exp \left( -\frac{1}{2} \| {\bf y}_n - {\bf H}_n^{k-} \hat{\bf m}_{n|n}^{k-,i} - {\bf H}_n^k {\bf x}_n^k\|_{{\bf R}_n^{-1}}^{2} \right), \\ 
\nonumber 
& = &  (\ref{eq-trans-y-N-v0-vb}). 
\end{eqnarray}

We now derive ${\pi}_{n-1}^i({\bf x}_{n}^k)$  (\ref{vb-trans-pdf-xk})-(\ref{vb-trans-pdf-xk-cov}). One first shows that, 
\begin{eqnarray}  
\nonumber 
{\pi}_{n-1}^i({\bf x}_{n}^k)  & \propto  &    \exp  \left(\mathbb{E}_{\pi_n^{(i,i-1)}({\bf x}_{n}^{k^-})} [\ln p_{n-1}({\bf x}_{n})] \right), \\   
\label{vb-prior-theo-app} 
  & \propto  &  \exp  \left( \! \mathbb{E}_{\pi_n^{(i,i-1)}({\bf x}_{n}^{k^-})} [\ln p({\bf x}_n^k|{\bf x}_n^{k-}, {\cal Y}_{n-1})] \! \right). 
\end{eqnarray} 
 On the other hand, the Gaussian assumption on  $p_{n-1}({\bf x}_n)$ (i.e., ${\cal N}_{{\bf x}_n} (\hat{\bf x}_{n|n-1} , {\bf P}_{n|n-1})$) implies that the  conditional distribution $p({\bf x}_n^k|{\bf x}_n^{k-}, {\cal Y}_{n-1})$ is also Gaussian with mean $\boldsymbol{\mu}_n({\bf x}_n^{k-}) =  \hat{\bf x}_{n|n-1}^k + {\bf G}_n^k ({\bf x}_{n}^{k-} - \hat{\bf x}_{n|n-1}^{k-})$, and covariance $\boldsymbol{\Sigma}_n = {\bf P}_{n|n-1}^k - {\bf G}_n^k {\bf P}_{n|n-1}^{k-,k}$, where    ${\bf G}_n^k = {\bf P}_{n|n-1}^{k,k-} ({\bf P}_{n|n-1}^{k-})^{-1}$, and  ${\bf P}_{n|n-1}^{k}$, ${\bf P}_{n|n-1}^{k-}$, ${\bf P}_{n|n-1}^{k,k-}$ and ${\bf P}_{n|n-1}^{k-,k}$ are extracted from ${\bf P}_{n|n-1}$ according to the  positions $k$ and $k^-$ \cite[e.g.,][Appendix, Prop.  $9$]{aitelfquih-et-desbouvries-jsps-2011}. In an ensemble setting, the supposedly large $(d_x - d_p) \times (d_x - d_p)$-sized covariances ${\bf P}_{n|n-1}^{k-}$ are  approximated based on a small number $M$ of forecast  members, meaning that these  are singular, thus non-invertible. In such a case, the inverse $({\bf P}_{n|n-1}^{k-})^{-1}$ in the expression of ${\bf G}_n^k$ is replaced by a pseudo-inverse, $({\bf P}_{n|n-1}^{k-})^{+}$ \cite[e.g.,][Sec. 2.3]{andersonmoore}.  \newline 
 Now, following  a similar  derivation to that of ${\cal L}_{{\bf y}_n}^i({\bf x}_{n}^k)$ above, we apply the $\ln$ function to $p({\bf x}_n^k|{\bf x}_n^{k-}, {\cal Y}_{n-1}) = {\cal N}_{{\bf x}_n^k}(\boldsymbol{\mu}_n , \boldsymbol\Sigma_n)$, followed by the  expectation w.r.t.  $\pi_n^{(i,i-1)}({\bf x}_{n}^{k^-})$, then the $\exp$ function, to  readily obtain (\ref{vb-trans-pdf-xk})-(\ref{vb-trans-pdf-xk-cov}).  
 
 One still has to derive ${\pi}_{n}^i({\bf x}_{n}^k)$ (\ref{vb-trans-pdf-xk-analysis})-(\ref{vb-trans-pdf-xk-cov-analysis}). Since the VB pdf of ${\bf x}_n^k$, ${\pi}_{n-1}^i({\bf x}_{n}^k)$ (\ref{vb-trans-pdf-xk}), is Gaussian, and the VB pdf of ${\bf y}_n$ conditioned on ${\bf x}_n^k$, ${\cal L}_{{\bf y}_n}^i({\bf x}_{n}^k)$ (\ref{eq-trans-y-N-v0-vb}), is Gaussian with a mean linear in ${\bf x}_n^k$, then the associated joint pdf, $\kappa^i({\bf x}_n^k, {\bf y}_n) = {\pi}_{n-1}^i({\bf x}_{n}^k) {\cal L}_{{\bf y}_n}^i({\bf x}_{n}^k)$, is Gaussian with parameters given as \cite[e.g.,][Appendix, Prop.  $8$]{aitelfquih-et-desbouvries-jsps-2011},  
 \begin{eqnarray} 
 \label{eq-joint-vb-pdf}  
\kappa^{i}({\bf x}_n^k, {\bf y}_n) \!\!\!\!\! & = \!\!\!\!\! & {\cal N}_{({\bf x}_n^k, {\bf y}_n)} \!  \left( 
\! \left[ \!\!\!
\begin{array}{c} 
{\bf m}_{n|n-1}^{k,i} \\ 
{\bf H}_{n}^k {\bf m}_{n|n-1}^{k,i} + {\bf H}_n^{k-} {\bf m}_{n|n}^{k-,i}
\end{array}  
\!\!\! \right]   \! , \! \left[ \!\!\! \begin{array}{cc} 
{\bf C}_{n|n-1}^k \! & \! {\bf C}_{n|n-1}^k ({\bf H}_n^k)^T \\ 
{\bf H}_n^k {\bf C}_{n|n-1}^k \! & \! {\bf H}_n^k{\bf C}_{n|n-1}^k ({\bf H}_n^k)^T + {\bf R}_n 
\end{array}
\!\!\! \right] \!
\right) 
\end{eqnarray}
%
  The VB pdf of interest, ${\pi}_{n}^i({\bf x}_{n}^k) = \kappa^{i}({\bf x}_n^k| {\bf y}_n)$, is then obtained from a conditioning of Eq. (\ref{eq-joint-vb-pdf}) using Prop. $9$ of \citep[][Appendix]{aitelfquih-et-desbouvries-jsps-2011}. By doing so, one immediately obtains (\ref{vb-trans-pdf-xk-analysis})-(\ref{vb-trans-pdf-xk-cov-analysis}).

\vspace{.05cm}  
\begin{center}
\textbf{The practical computation of NFE over iterations} 
\end{center} 

From Eq. (\ref{eq-NFE-max}), the NFE  at assimilation time $t_n$ and iteration $i$ is defined as,   
\begin{eqnarray}
\nonumber  
{\rm NFE}^i  & =  &   \mathbb{E}_{\pi_n^i({\bf x}_n)} \left[ \ln p({\bf x}_n, {\bf y}_n | {\cal Y}_{n-1}) \right]  -   
  \overbrace{ \mathbb{E}_{\pi_n^i({\bf x}_n)} \left[ \ln \pi_n^i({\bf x}_n) \right] }^{\phi_3^i} \\ 
\label{eq_nfe-at-iteration-i}
 & =  & \underbrace{ \mathbb{E}_{\pi_n^i({\bf x}_n \!)}  \left[ \ln p({\bf y}_n | {\bf x}_{n}) \right] }_{\phi_1^i } +     \underbrace{ \mathbb{E}_{\pi_n^i({\bf x}_n \!)}  \left[ \ln p_{n-1}({\bf x}_n) \right] }_{\phi_2^i}  -    \phi_3^i .
\end{eqnarray} 
Inserting in the above $\phi^i$ terms the Gaussians $p({\bf y}_n | {\bf x}_{n})  =  {\cal N}_{{\bf y}_n} ({\bf H}_n {\bf x}_n , {\bf R}_n)$,  
$p_{n-1}({\bf x}_n)   \approx  {\cal N}_{{\bf x}_n} (\hat{\bf f}_{n} , {\bf P}_{{\bf f}_n})$ and  $\pi_{n}^i({\bf x}_n)  \approx  {\cal N}_{{\bf x}_n} (\hat{\bf a}_{n}^i , {\bf P}_{{\bf a}_n})$, with ${\bf P}_{{\bf f}_n}$ and ${\bf P}_{{\bf a}_n}$ are block-diagonal covariances formed based on the ensemble perturbation matrices ${ \{ {\bf S}_{{\bf f}_n^k} \} }_k$ and  ${ \{ {\bf S}_{{\bf a}_n^k} \} }_k$, one obtains, 
  \begin{equation}  
 \label{eq-nfe-phiTerms}
\left\{ \!  
\begin{array}{ccl}   
	\phi_1^i     & \approx   &  - \frac{1}{2}  \left[ \ln ((2\pi)^{d_y} |{\bf R}_n|) + 
	    {\rm Tr}( {\bf P}_{{\bf a}_n} {\bf H}_n^T {\bf R}_n^{-1} {\bf H}_n)   +  ||{\bf y}_n - {\bf H}_n \hat{\bf a}_{n}^i ||_{{\bf R}_n^{-1}}^2   \right] , \\ 
    \phi_2^i   & \approx   &  -\frac{1}{2}   \left[ \ln ((2\pi)^{d_x} |{\bf P}_{{\bf f}_n}|)  +   {\rm Tr}({\bf P}_{{\bf a}_n} {\bf P}_{{\bf f}_n}^{-1})    +  || \hat{\bf a}_{n}^i  -  \hat{\bf f}_{n} ||_{{\bf P}_{{\bf f}_n}^{-1}}^2      \right] , \\ 
    \phi_3^i  \!\!  & \approx \!\!  &  -\frac{1}{2} \!  \left[ \ln ((2\pi)^{d_x} |{\bf P}_{{\bf a}_n}|) + d_x   \right] . 
\end{array} 
\right.  
\end{equation} 
 %
 %
 Eqs. (\ref{eq-nfe-phiTerms}) are then inserted in Eq. (\ref{eq_nfe-at-iteration-i}) to obtain an ensemble-based approximate expression of ${\rm NFE}^i$.

\section*{Acknowledgments}
The authors  thank the anonymous reviewers for  the valuable and  constructive comments.

\section*{Data Availability Statement} 

Data sharing not applicable – no new data generated.

\section*{Author Contributions}
Boujemaa Ait-El-Fquih: conceptualization; investigation; writing - original draft; methodology; writing - review \& editing.  \\[0.5em] 
Ibrahim Hoteit: conceptualization; funding acquisition; resources; writing - review \& editing.


%

\section*{ORCID}
Boujemaa Ait-El-Fquih  \orcidlink{0000-0003-1119-7585} \url{https://orcid.org/0000-0003-1119-7585} \\[0.5em]
Ibrahim Hoteit \orcidlink{0000-0002-3751-4393} \url{https://orcid.org/0000-0002-3751-4393}



\end{document}